\documentclass[final]{cvpr}

\usepackage{times}
\usepackage{epsfig}
\usepackage{graphicx}
\usepackage{amsmath}
\usepackage{amssymb}
\usepackage{subcaption}
\usepackage[super]{nth}
\usepackage{siunitx}
\usepackage{nopageno}

\usepackage[pagebackref=true,breaklinks=true,colorlinks,bookmarks=false]{hyperref}

\def\cvprPaperID{****} 
\def\confYear{CVPR 2021}

\begin{document}

\title{Attention! Stay Focus!}

\author{Tu Vo\\
BridgeAI Inc.\\
Seoul, South Korea\\
{\tt\small tuvovan@pukyong.ac.kr}
}

\maketitle

\begin{abstract}
   We develop a deep convolutional neural networks (CNNs) to deal with the blurry artifacts caused by the defocus of the camera using dual-pixel images. Specifically, we develop a double attention network which consists of attentional encoders, triple locals and global local modules to effectively extract useful information from each image in the dual-pixels and select the useful information from each image and synthesize the final output image. We demonstrate the effectiveness of the proposed deblurring algorithm in terms of both qualitative and quantitative aspects by evaluating on the test set in the NTIRE 2021 Defocus Deblurring using Dual-pixel Images Challenge~\cite{ntire}. The code, and trained models are available at  \url{https://github.com/tuvovan/ATTSF}.
\end{abstract}

\section{Introduction}
In general, the exposure of the image can be adjusted by two different ways. The first way is to change the shutter speed when the aperture is fixed to control the amount of light falling on the sensor. The other way is to keep the shutter speed unchanged while adjusting the aperture's size. While the former method may cause the motion blur if there is any possible object motion, the later method results in a shallow depth of field (DoF), causing defocus blur to occur in scene regions outside the DoF~\cite{abuolaim2020defocus} .
Removing the defocus blur is critical as we can obtain an image which is captured using a wide aperture but still have everything in focus, which ensures a well-exposed image with sufficiently sharp image.

In theory, defocus blur is a result of a sharp region with a spatial point spread function (PSF) that use the neighborhood pixel in producing the blurry pixel~\cite{tang}. As a result, using the dual-pixel alone may not sufficiently enough to faithfully recover the original sharp pixel. However, we believe that by employing the large receptive fields provided by stacking convolution layers with maxpooling, a neural network will be able to produce the non-blurred outputs, given the dual-pixel inputs.
Recently, Abuolaim \textit{et al.}~\cite{abuolaim2020defocus} trained an Unet-like model which takes two images as input and produces a defocus blur-free output. Unet is an encoder-decoder framework which considers all pixel and channel equally which we think not suitable as the blurry pixels are distributed differently for each channel and each pixel location.

In this work, to employ different feature in both input images intentionally, we propose an attention deep convolutional neural networks (CNNs) to remove the defocus blur artifact which is built upon Encoder - Decoder architecture with the Dual Attention Modules. As mentioned above, we notice that every pixel and channel of the input images should be considered appropriately, make them contribute to the final output at different level.  As a result, we redesigned the encoder to extract the useful information by adding the dual-attention module to the classical encoder module. Furthermore, the extracted features from the attention-encoder will be concatenated and put through the triple-local and global-non-local modules before being decoded by decoder modules and generate the sharp output image. We demonstrate the effectiveness of the proposed network through the \textit{NTIRE2021 Defocus Deblurring Challenge~\cite{ntire}}. Using the data provided by the competition, we trained a network and finally archived the average \textit{PSNR} of $26.4243$ \textit{dB}, stands at the \nth{9} position in the competition.

\section{Related Works}

\subsection{Defocus Deblurring}

While there are many previous works on deblurring field, we found that those methods that try to estimate the defocus map and deblurring are the closest methods as they all try to produce the sharp and deblurred output. The most common method for defocus deblurring task is to first estimate the deblurring kernel and then use that kernel as a guidance for deblurring. To find the deblurring kernel, Park \textit{et al.}~\cite{park} fed a combination of pretrained blur classification network to extract the deep blur feature along with the hand-crafted feature to a regression network to estimate the amount of blur in the pixel edge to later deblur it. Karaali \textit{et al.}~\cite{karaali} extracted the difference between gradient of the blurry image and the original one. And most recently, Abuolaim \textit{et al.}~\cite{abuolaim2020defocus}  introduced a deep learning model, which consists of encoder and decoder modules, use the dual-pixel data to directly solve for the defocus blur in a single step without estimating the defocus map.

\subsection{Attention Modules in Deep Learning}

Recently, attention mechanisms show the effectiveness in many computer vision fields including the image restoration. Thanks to its ability to facilitate deep neural networks to determine where to focus and improve the representation of interest~\cite{an}. By observing that each pixel and each image channel should be considered separately, we end up adding the dual-attention~\cite{dual} module to the conventional encoder in our proposed network to tackle with the defocus deblurring challenge. 
By incorporating the attention mechanisms with the conventional encoder modules, every pixel and channel is dealt separately, make sure they contribute the useful information at different level before being merged and being decoded. With this mechanism, the proposed architecture yields high-quality results in both qualitative and quantitative perspective.

\subsection{Defocus Blur Dataset}

There are several available datasets for the defocus deblurring task. Salvado \textit{et al.}~\cite{salvado} proposed the RTF dataset which has 22 image pairs of blurred image and its corresponding in-focus-image. The CUHK~\cite{shi} and DUT~\cite{zhao} provide real-blurred images with the corresponding binary masks representing the blur/sharp regions. This dataset is more suitable for blur detection task. Recently, Abuolaim \textit{et al.}~\cite{abuolaim2020defocus}  proposed a large dataset with 500 different pairs of non-overlapped scenes. By using the dual-pixels, the dataset is extended to 2000 images with blur images and the corresponding sharp images. This dataset is used for \textit{NTIRE2021 Defocus Deblurring Challenge~\cite{ntire}}.

\begin{figure*}
\begin{center}
\includegraphics[scale=0.43]{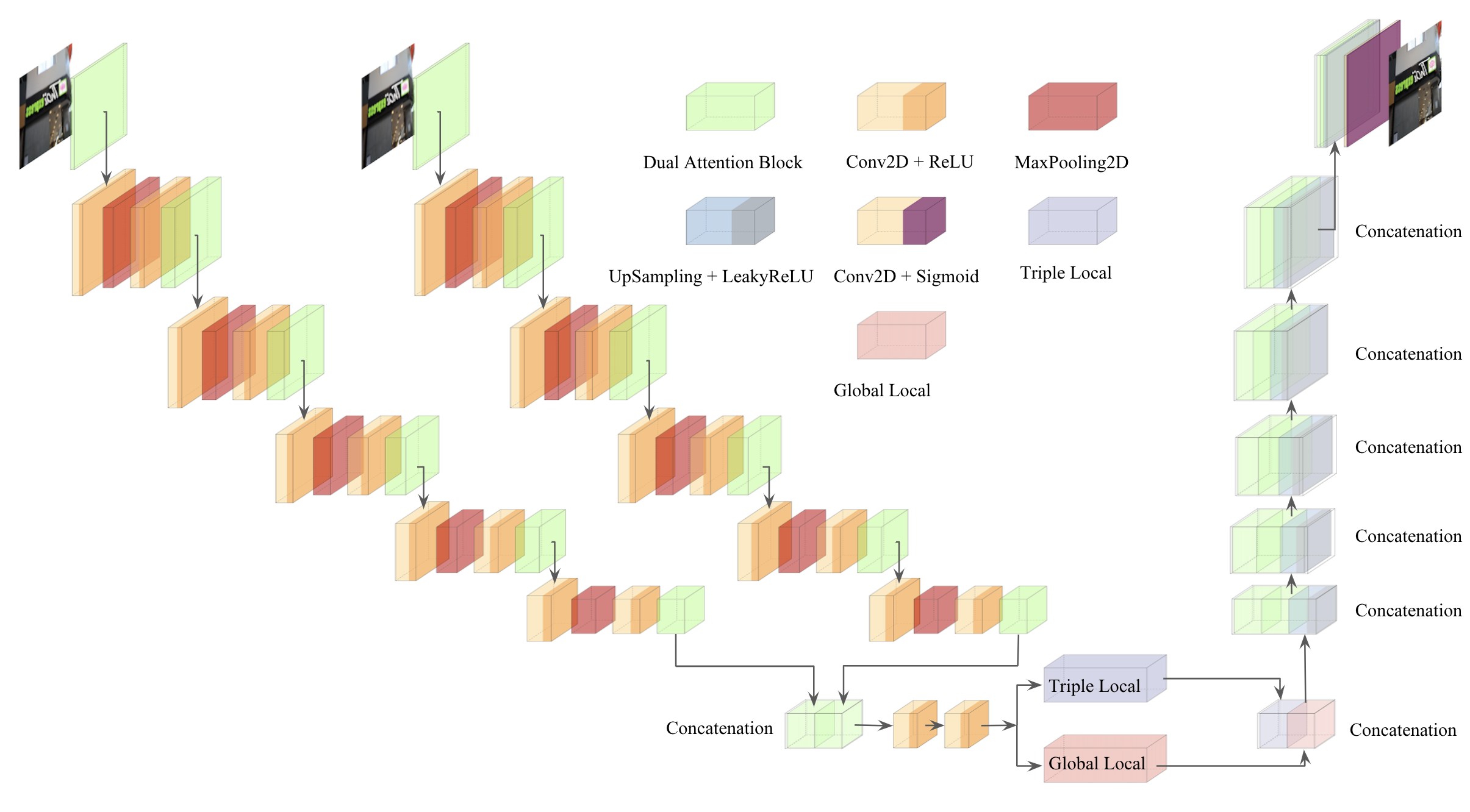}
\end{center}
   \caption{The overall architecture of the proposed ATTSF.}
\label{fig:overall}
\end{figure*}

\section{Attention! Stay Focus! (ATTSF)}
We design an attention encoder decoder network to effectively synthesize the blurry input images and generate a high-quality blur-free output. Figure~\ref{fig:overall} shows the architecture of the proposed network, which takes two images (left blurry image and right blurry image) as input and then reconstructs a sharp image. In details,  our proposed ATTSF consists of several attention encoders, triple local, global-local blocks and decoder modules. The attention encoders are used to extract useful features from the blurry input images. The features generated from those encoders are concatenated together and being transferred to the triple-local and global non-local modules in parallel then finally being decoded to get the final sharp output image. To ensure the output image to have the useful feature from the input images, we use the skip-connection to connect the output feature of the encoders and decoders at every level.

\subsection{Attention Encoder (ATTE)}

\begin{figure*}
\begin{center}
\includegraphics[scale=0.45]{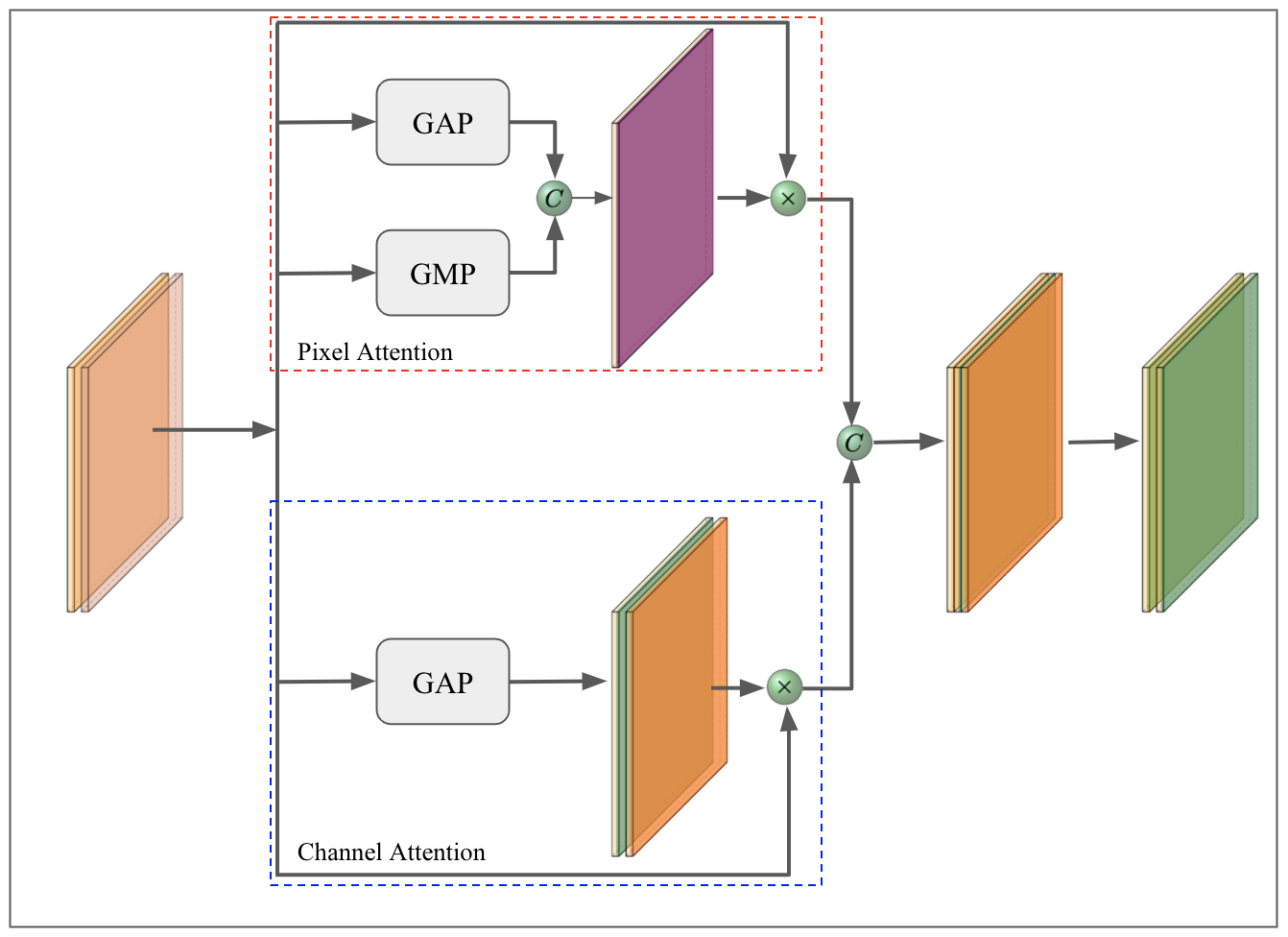}
\end{center}
   \caption{Dual Attention Module. GAP, GMP are Global Average Pooling and Global Max Pooling, respectively. $\times$ denotes channel-wise multiplication, and $C$ denotes the concatenate operation.}
\label{fig:dual}
\end{figure*}

The conventional encoder usually consists of several convolution layers following by the activation layers and pooling layers. Encoder modules are good at extracting the high-level feature of the input image. However, all pixel and channel are treated equivalently which is not strong enough in this defocus deblurring challenge in our opinion. We observe that the defocus deblurring challenge, the blur level are not equally distributed both over image channels and image pixel. As a results, we employ the dual-attention mechanism, which is composed of channel attention and pixel attention or position attention. Figure~\ref{fig:dual} shows the architecture of the dual-attention modules. To be specific, each ATTE block has its input to go through the dual-attention module to extract the high-level feature intentionally. Follow the dual-attention module is a couple of convolution layers with ReLU~\cite{relu} activation functions and MaxPooling layers, just similar to the conventional encoder.

\textbf{Channel Attention}

As the input of the network is dual-pixel, each image should contribute different kind of information to the final output image. Base on this we employed the dual-attention module which includes the channel attention and pixel attention. The channel attention intentionally extract the feature across the channel dimension by calculating the channel attention map from the input feature. The channel attention applies the convolution layer following by the sigmoid function, which ensures that the attention map will range from 0.0 to 1.0, representing the amount of the information contribution of each feature channel to the output. By masking the input feature with the calculated attention mask,  we get the output feature which consists of useful information from each channel of the input.

\textbf{Pixel Attention}

In addition to channel attention, pixel attention a.k.a position attention is also crucial for this task. While channel attention works on channel domain, pixel attention, on the other hand, pays attention to the every pixel in the input feature. The module applies global average pooling(GAP) and global max pooling(GMP) in parallel on the input feature. The GAP and GMP feature are then concatenated together, follows by a $1\times1$ convolution with the sigmoid activation function to generate a pixel attention mask. The attention mask is then multiply equally with every input channel, resulting in the output feature.

\textbf{Dual Attention}

Having the pixel attention and channel attention, the dual attention module takes the input feature and applies two $3\times3$ convolution layers follows by ReLU~\cite{relu} activation function. The feature is then put through Pixel Attention and Channel Attention simultaneously and being concatenated in the channel axis. Finally, the concatenated feature is $1\times1$ convoluted to match the dimension of the input feature, as shown in Figure~\ref{fig:dual}.

\textbf{Attention Encoder}

As mentioned before, the proposed attention encoder is built based on the conventional encoder scheme by adding the dual attention module on top of it. Specifically, the input feature first goes through the dual attention module, then through the encoder part, which is composed of several $3\times3$ convolution layers and ReLU activation functions~\cite{relu}.

\begin{figure*}
\begin{center}
\includegraphics[scale=0.45]{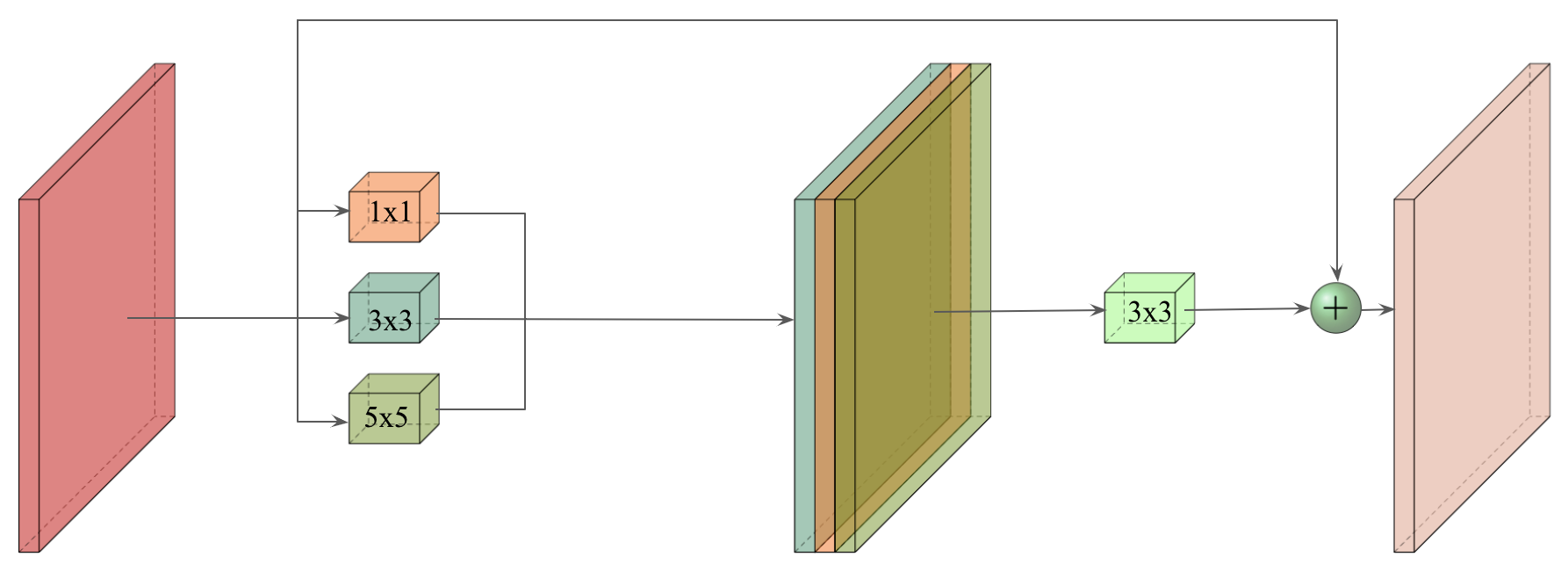}
\end{center}
   \caption{Triple Local Module}
\label{fig:triple}
\end{figure*}

\subsection{Triple Local}
Figure~\ref{fig:triple} shows the architecture of the triple local module. This is inspired by the inception modules, which has multiple convolution kernel with different size, in order to extract the feature of different levels. The small filter is able to extract local details of the features, and the large filter can cover larger regions of the receiving layers. All the features are concatenated in a channel-wise manner and compressed through a convolutional layer.

\begin{figure*}
\begin{center}
\includegraphics[scale=0.55]{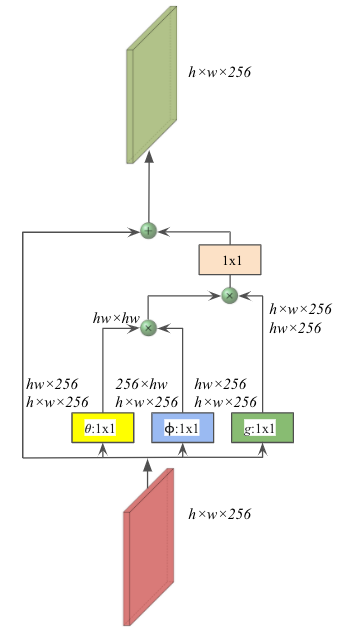}
\end{center}
   \caption{Global Local Module}
\label{fig:short}
\end{figure*}

\subsection{Global Local}
Figure 4 illustrates the architecture of the global local module. As we know, the convolution represents the local feature. In this task, although the local features are essential, we do not want to loose the global terms as it makes the whole image to be spatially consistent. Here we employed the idea from~\cite{wang} and~\cite{buades} which calculate the correlation between two input signals of the whole image. The global local module cover large receptive fields so the network can ensure the spatial consistency, avoiding the hallucination.

\subsection{Implementation Details}
In our implementation, each convolution in encoder module is followed by a Rectifier Linear Unit (ReLU) activation function~\cite{relu}, while each layer in decoder module is followed by a Leaky Rectifier Linear Unit (Leaky ReLU)~\cite{leakyrelu}. The reason behind using the Leaky ReLU instead of ReLU is to avoid the under-bound of the hidden layer's output, which may lead to unwanted reconstructed image. Every layer is initialized follow the \textit{He normal}~\cite{henormal}, and all convolution kernel in encoders or decoders are $3\times3$ . In each training batch, we apply several augmentation technique such as random rotation, horizontal and vertical flipping. All input images are normalized between 0.0 and 1.0. 

We first trained the networks using the Adam optimizer~\cite{adam} with the learning rate of \num{1e-4}, and the batch size was set to 4 for 200 epochs. We then change the loss function to loss function 2 as shown in Eq.~\ref{eqn:loss} and train the model with SGD optimizer with the batch size of 2, learning rate of \num{5e-5} with 100 more epochs, where $\alpha=1$, $\beta=0.5$, respectively. We also apply the Learning Rate Scheduler to decrease the learning rate by half every 20 epochs. Finally, the model is implemented using Tensorflow~\cite{tf} with the help of Tensorflow Addons package. We experimentally found that finetuning the network again with the loss function~\ref{eqn:loss} that has more weight on \textit{SSIM} , the model not only gets a high \textit{PSNR} but also high \textit{SSIM} value. And by achieving high \textit{SSIM} score, the final predicted images are more similar to the ground-truths, make them become closer to the real sharp images.

We used the training dataset provided by the \textit{NTIRE2021 Defocus Deblurring Challenge~\cite{ntire}}. The dataset is divided into three parts: training, validation and testing . We used the training set for training and validation set, test set for validating and testing the model, respectively. Because of the memory limitation, we did not use the original images for training, but we cropped into many patches of size $560\times560$ from the training and validation sets by sliding over images with strides of $140\times140$. As a results, we end up with more than $2000$ images for training and $500$ images for validation. For testing, we keep the original sizes. The training took approximately three days using a computer with Intel® Core™ i7, 32GB RAM, and Nvidia V100 GPU. After training and fine-tuning, we use the original test set provided by the competition. The model takes about $0.5$ second for one image, which is reasonably fast.

\begin{equation}
\label{eqn:loss}
  Loss = \alpha\times SSIMLoss + \beta\times MAELoss
\end{equation}

\section{Experimental Results}
\subsection{Quantitative and Qualitative Evaluation}

\begin{table}
\begin{center}
\begin{tabular}{|l|c|c|c|}
\hline
Method & \textit{PSNR} & \textit{SSIM}  & \textit{MAE} \\
\hline\hline
Abuolaim \textit{et al.}~\cite{abuolaim2020defocus}  & 25.13 & 78.59  & 0.0406 \\
Ours &\textbf{25.98} & \textbf{81.15} & \textbf{0.0377} \\
\hline
\end{tabular}
\end{center}
\caption{\textit{PSNR}, \textit{SSIM} and \textit{MAE} values of our proposed algorithm, compared with Abuolaim \textit{et al.}~\cite{abuolaim2020defocus}  The bold values indicates the better results. We can notice that our network is far better than the state of the arts.}
\label{table:kysymys}
\end{table}

We evaluate the performance of the proposed network using the test set provided by the \textit{NTIRE2021 Defocus Deblurring Challenge~\cite{ntire}}. The competition provides two different test sets. One of them has ground truth images which we use to compare the qualitative metrics with state-of-the-art methods recently, the other one does not include the ground truth, so we only use them for visual comparison, as shown on Figure~\ref{fig:three}.

Table \ref{table:kysymys} compares the \textit{PSNR}, \textit{MAE} and \textit{SSIM} calculated using the former test set mention above. We only compare with Abuolaim \textit{et al.}~\cite{abuolaim2020defocus}  method as we notice that Abuolaim \textit{et al.}~\cite{abuolaim2020defocus} 's method is the only method that tries to solve the defocus deblurring problem using deep learning model, which is close to our work. The proposed method outperforms the state-of-the-art algorithm in terms of qualitative metrics, as the feature of the input images are extracted attentionally, and contributed to output the sharp images.

\begin{figure*}
	\label{fig1}
        \begin{subfigure}[b]{0.245\textwidth}
                \includegraphics[width=\linewidth]{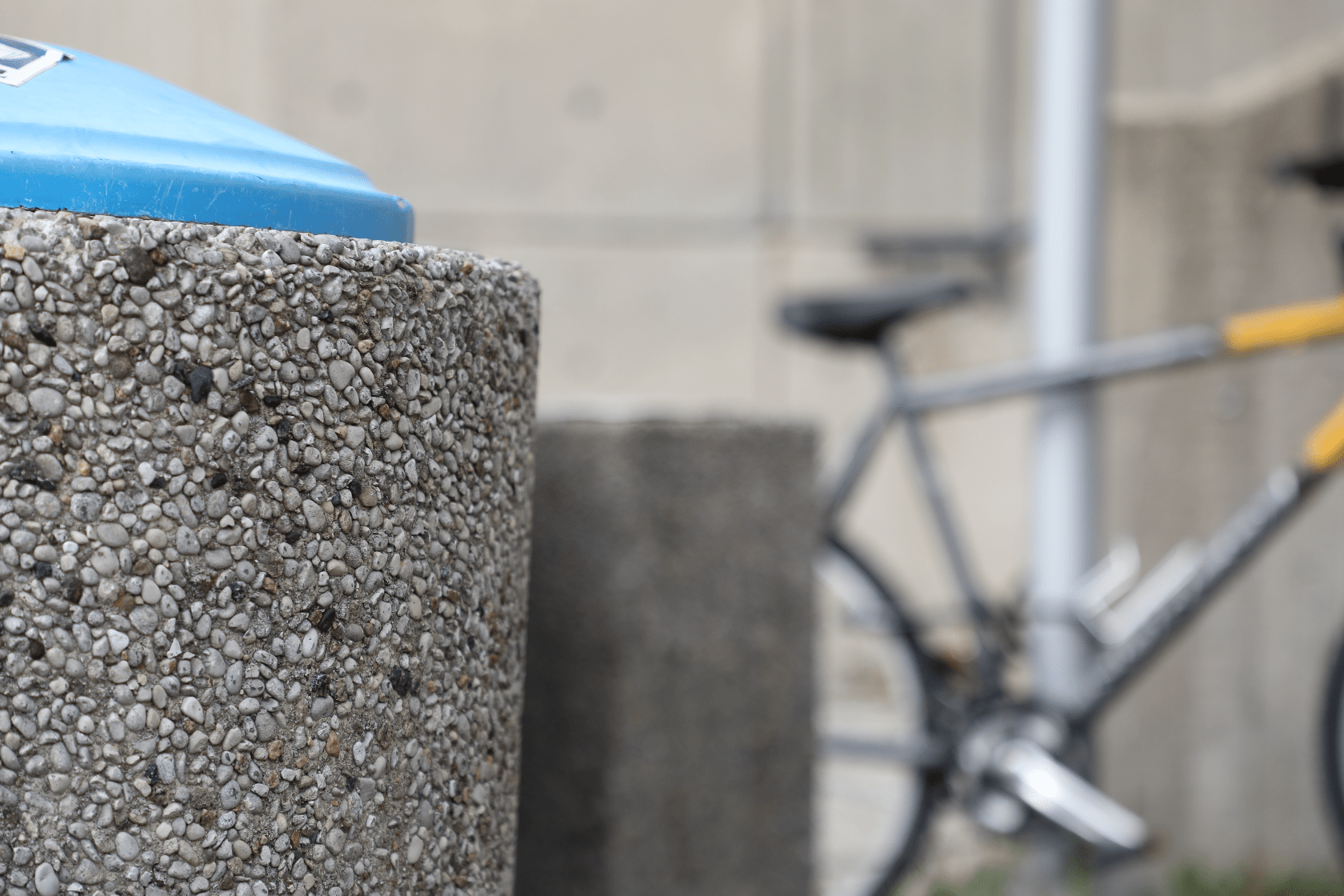}
        \end{subfigure}%
        \hspace{\fill}
        \begin{subfigure}[b]{0.245\textwidth}
                \includegraphics[width=\linewidth]{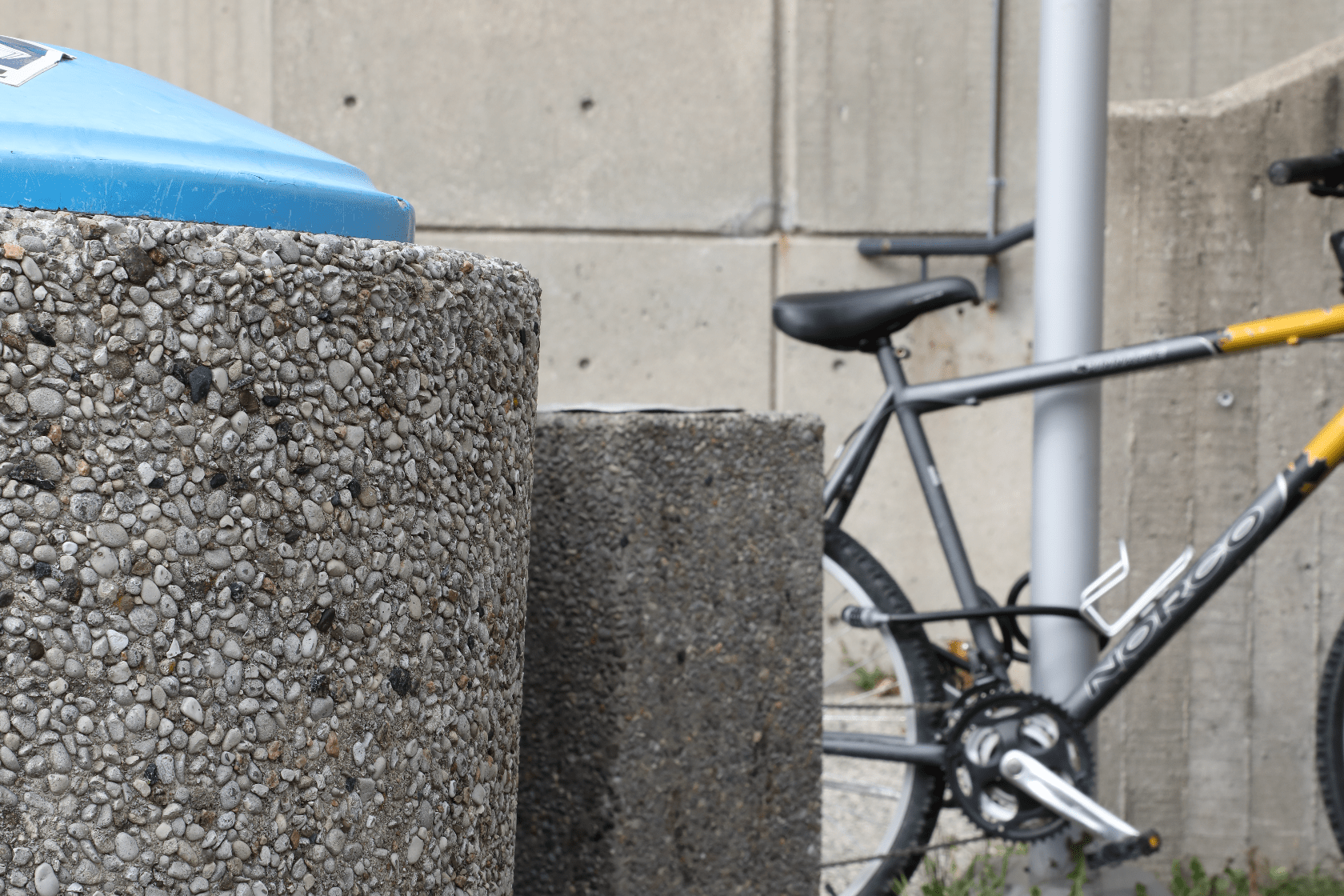}
        \end{subfigure}%
        \hspace{\fill}
        \begin{subfigure}[b]{0.245\textwidth}
                \includegraphics[width=\linewidth]{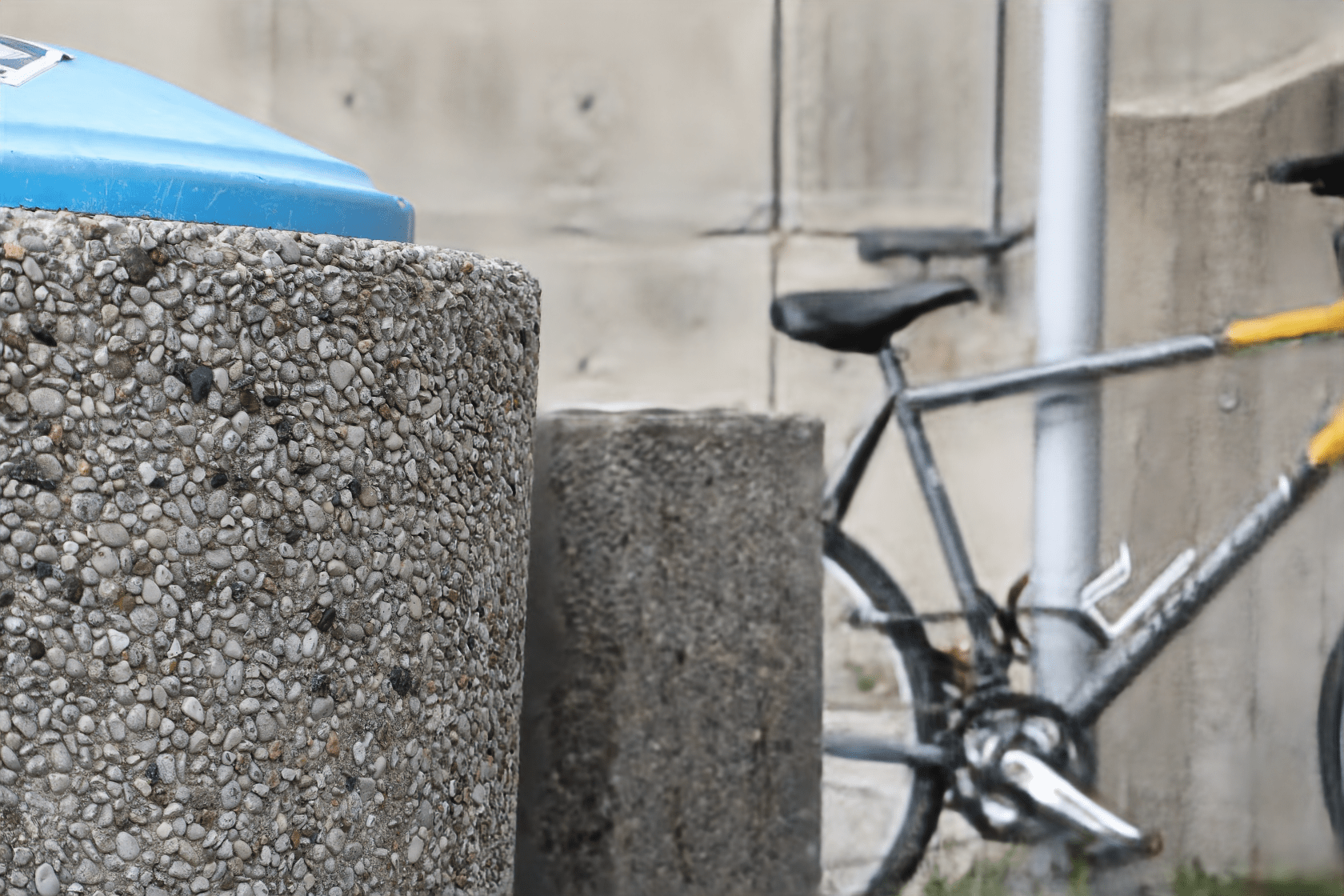}
        \end{subfigure}%
        \hspace{\fill}
        \begin{subfigure}[b]{0.245\textwidth}
                \includegraphics[width=\linewidth]{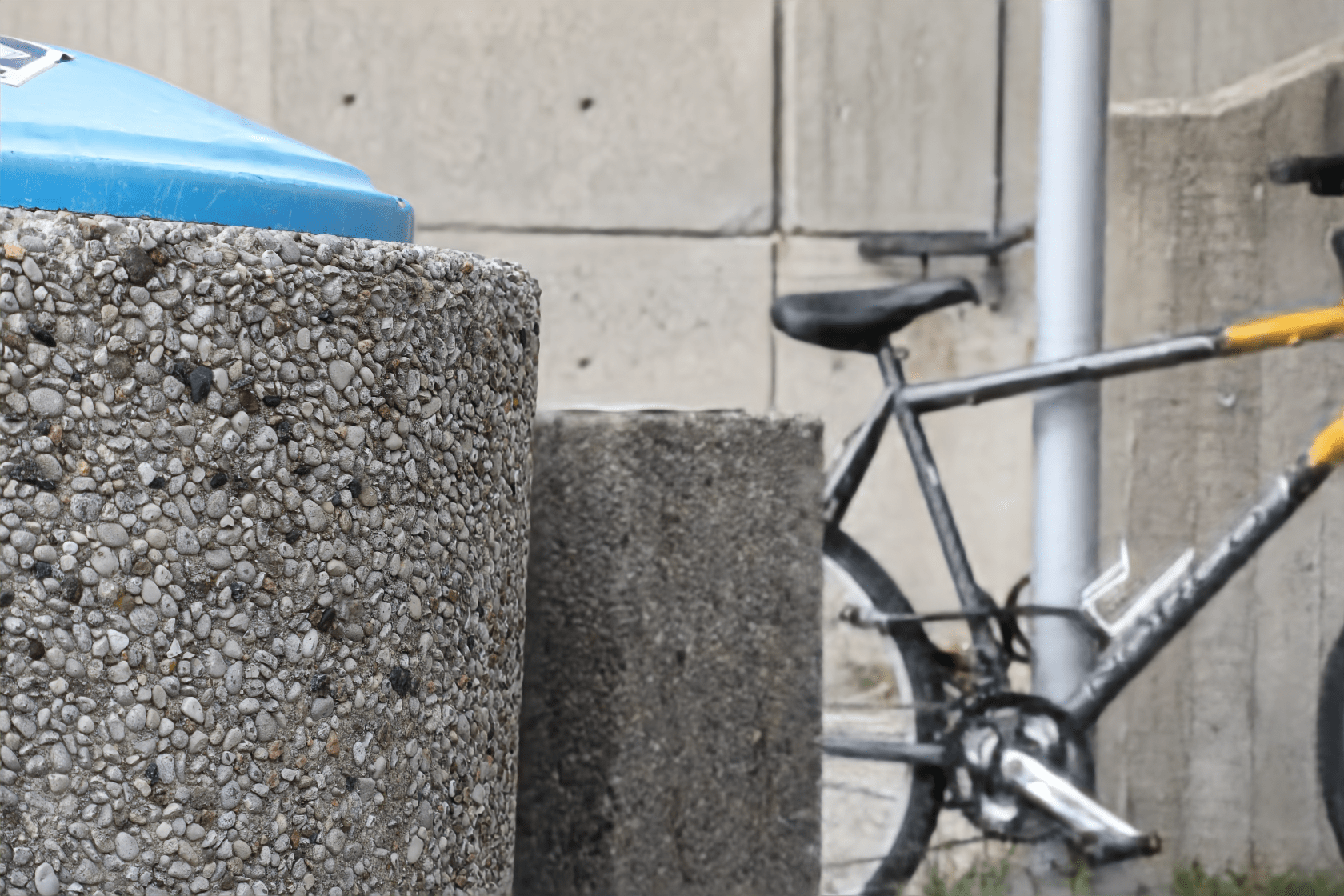}
        \end{subfigure}

        \begin{subfigure}[b]{0.245\textwidth}
                \includegraphics[width=\linewidth]{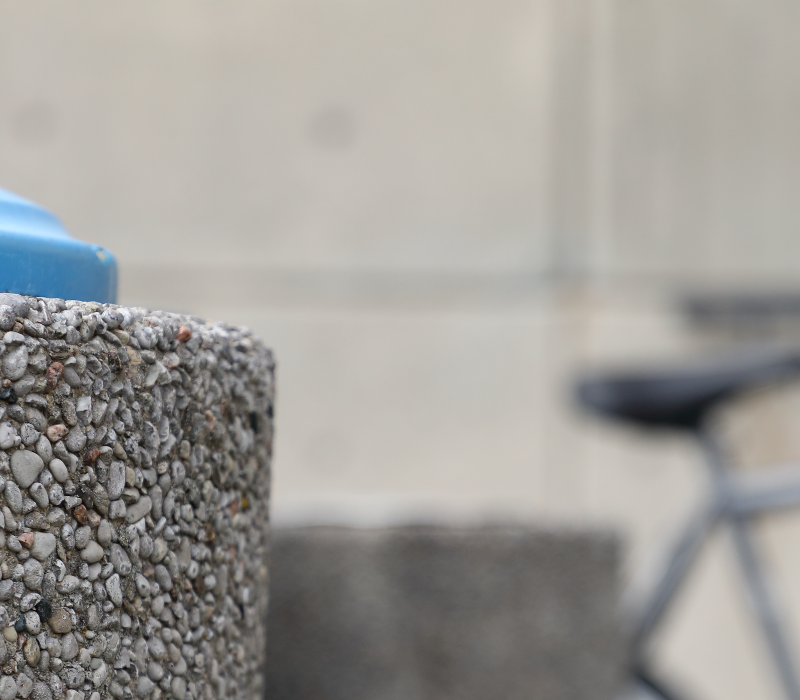}
                \caption{Input}
                \label{fig:gull}
        \end{subfigure}%
        \hspace{\fill}
        \begin{subfigure}[b]{0.245\textwidth}
                \includegraphics[width=\linewidth]{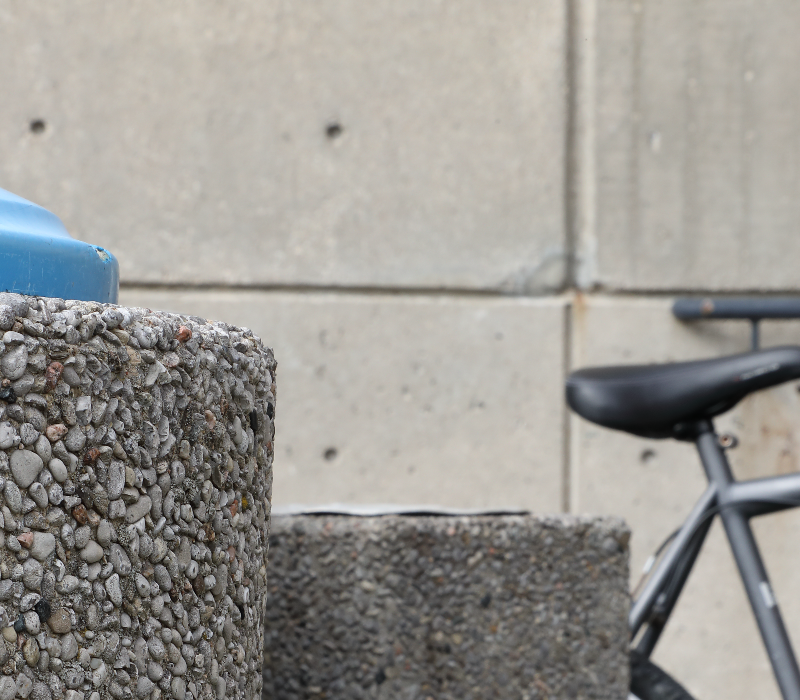}
                \caption{Ground Truth}
                \label{fig:gull2}
        \end{subfigure}%
        \hspace{\fill}
        \begin{subfigure}[b]{0.245\textwidth}
                \includegraphics[width=\linewidth]{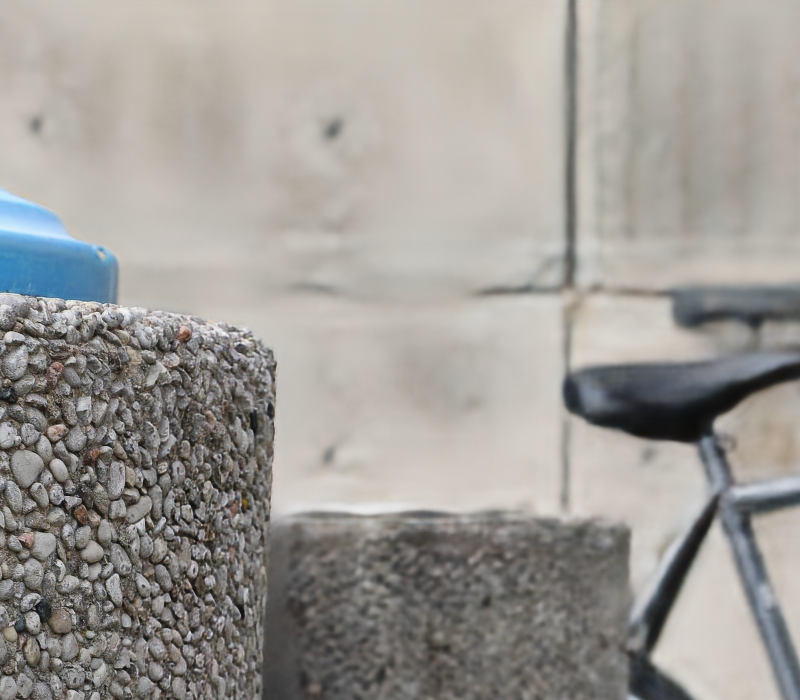}
                \caption{Abuolaim \textit{et al.}~\cite{abuolaim2020defocus} }
                \label{fig:tiger}
        \end{subfigure}%
        \hspace{\fill}
        \begin{subfigure}[b]{0.245\textwidth}
                \includegraphics[width=\linewidth]{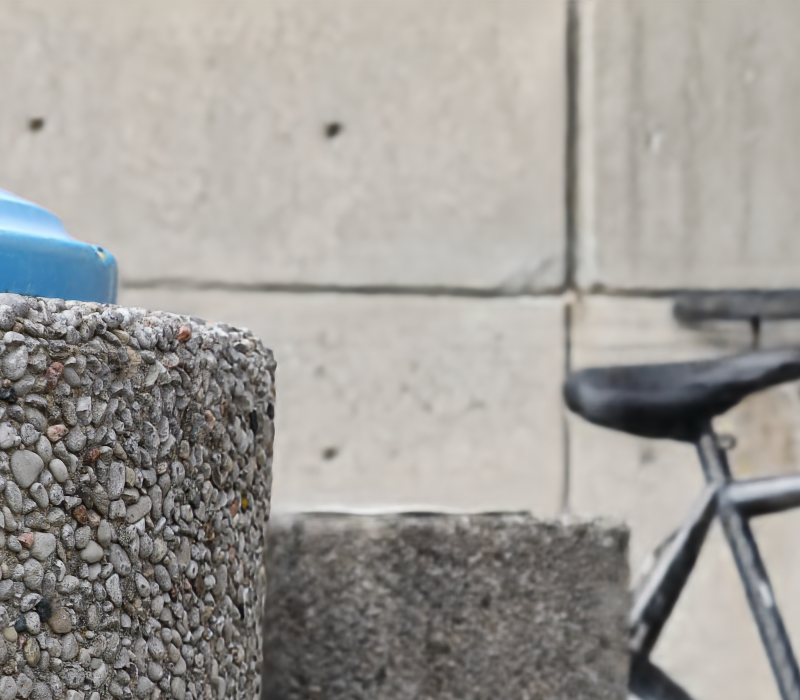}
                \caption{Proposed}
                \label{fig:mouse}
        \end{subfigure}
        \caption{Visual comparison of the proposed algorithm and Abuolaim \textit{et al.}~\cite{abuolaim2020defocus}  's algorithm. The proposed algorithm outperforms the Abuolaim \textit{et al.}~\cite{abuolaim2020defocus}  as it faithfully recovers the wall region, makes it close to the ground truth image.}\label{fig:one}
\end{figure*}

\begin{figure*}
	\label{fig2}
	\begin{subfigure}[b]{0.245\textwidth}
                \includegraphics[width=\linewidth]{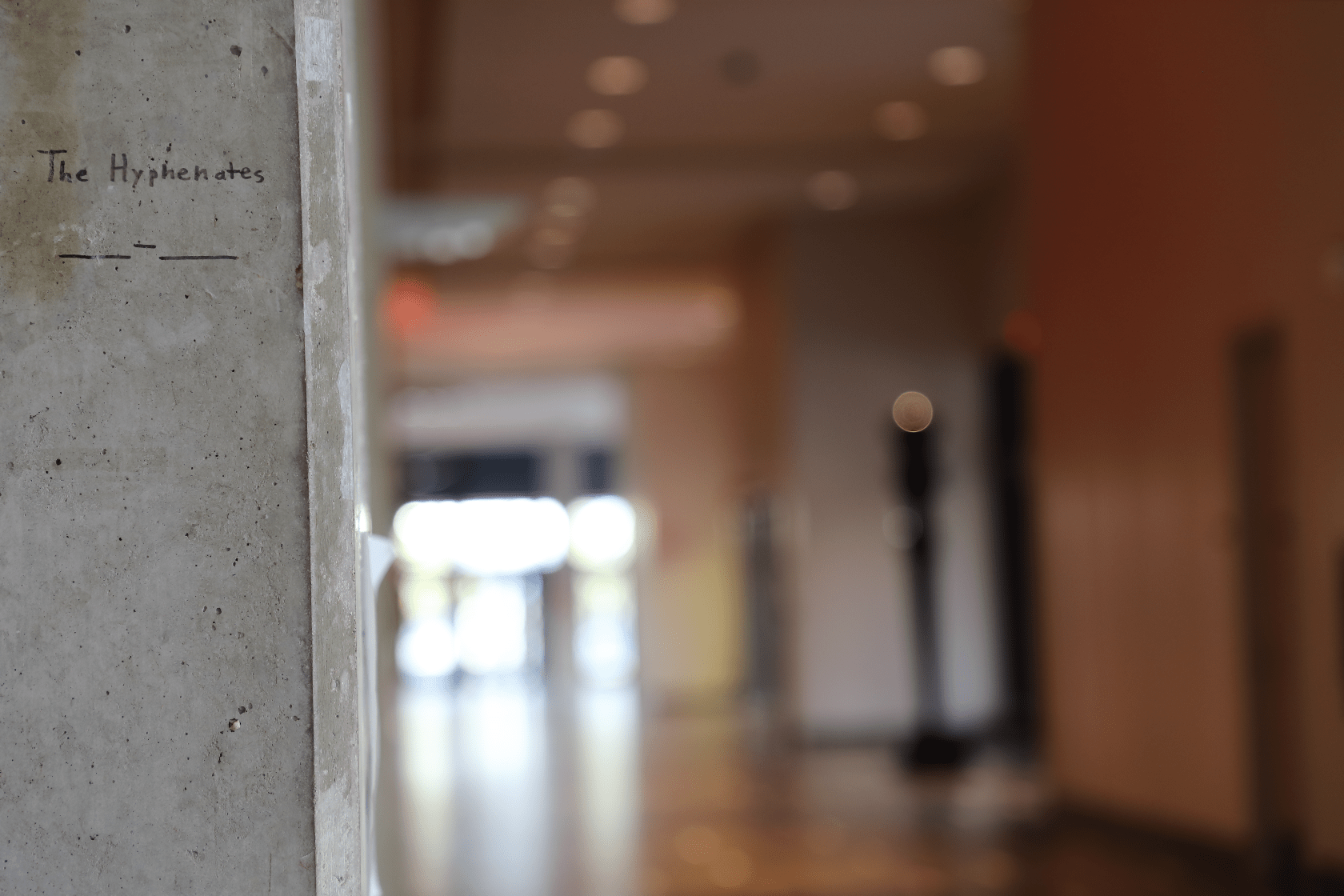}
        \end{subfigure}%
        \hspace{\fill}
        \begin{subfigure}[b]{0.245\textwidth}
                \includegraphics[width=\linewidth]{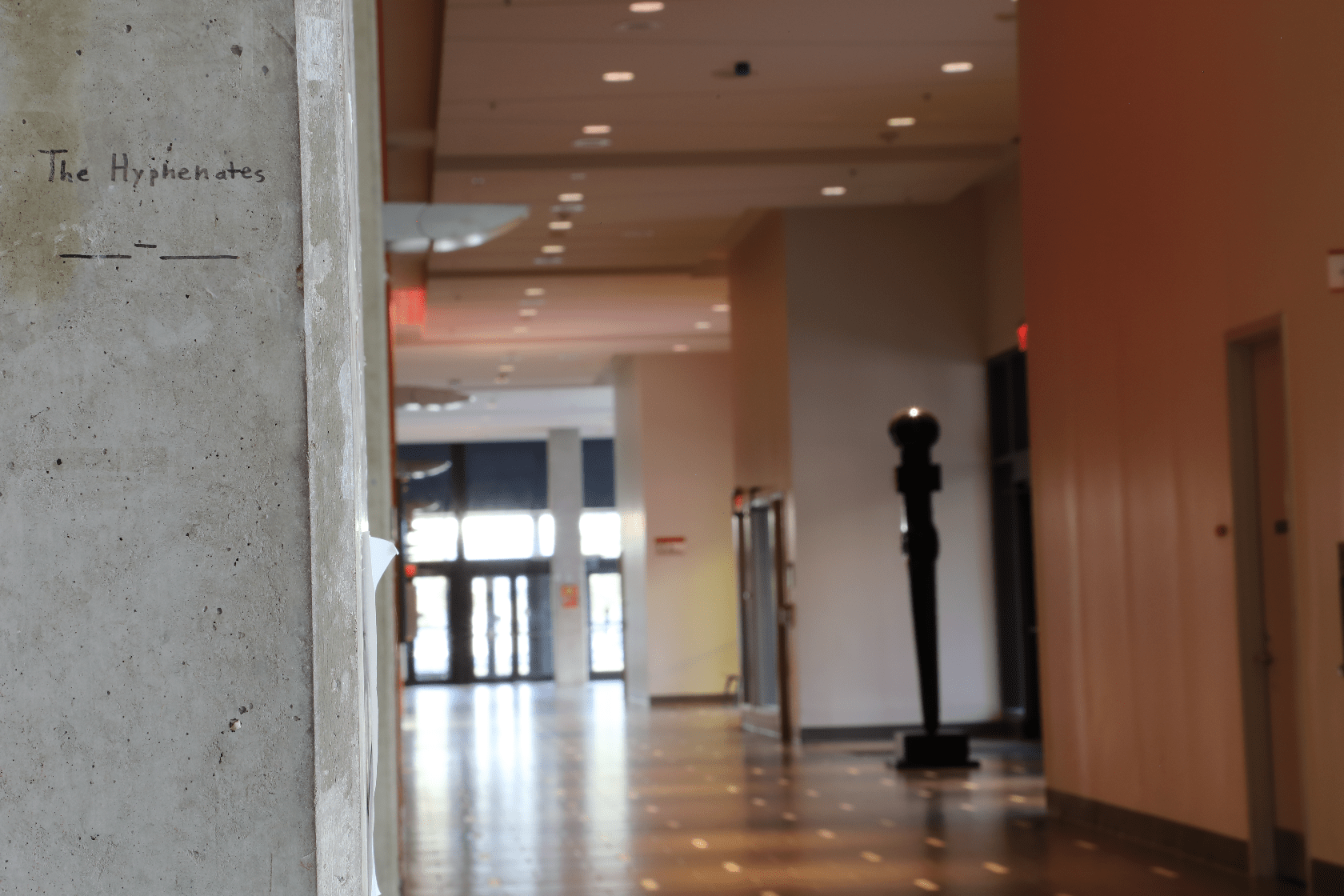}
        \end{subfigure}%
        \hspace{\fill}
        \begin{subfigure}[b]{0.245\textwidth}
                \includegraphics[width=\linewidth]{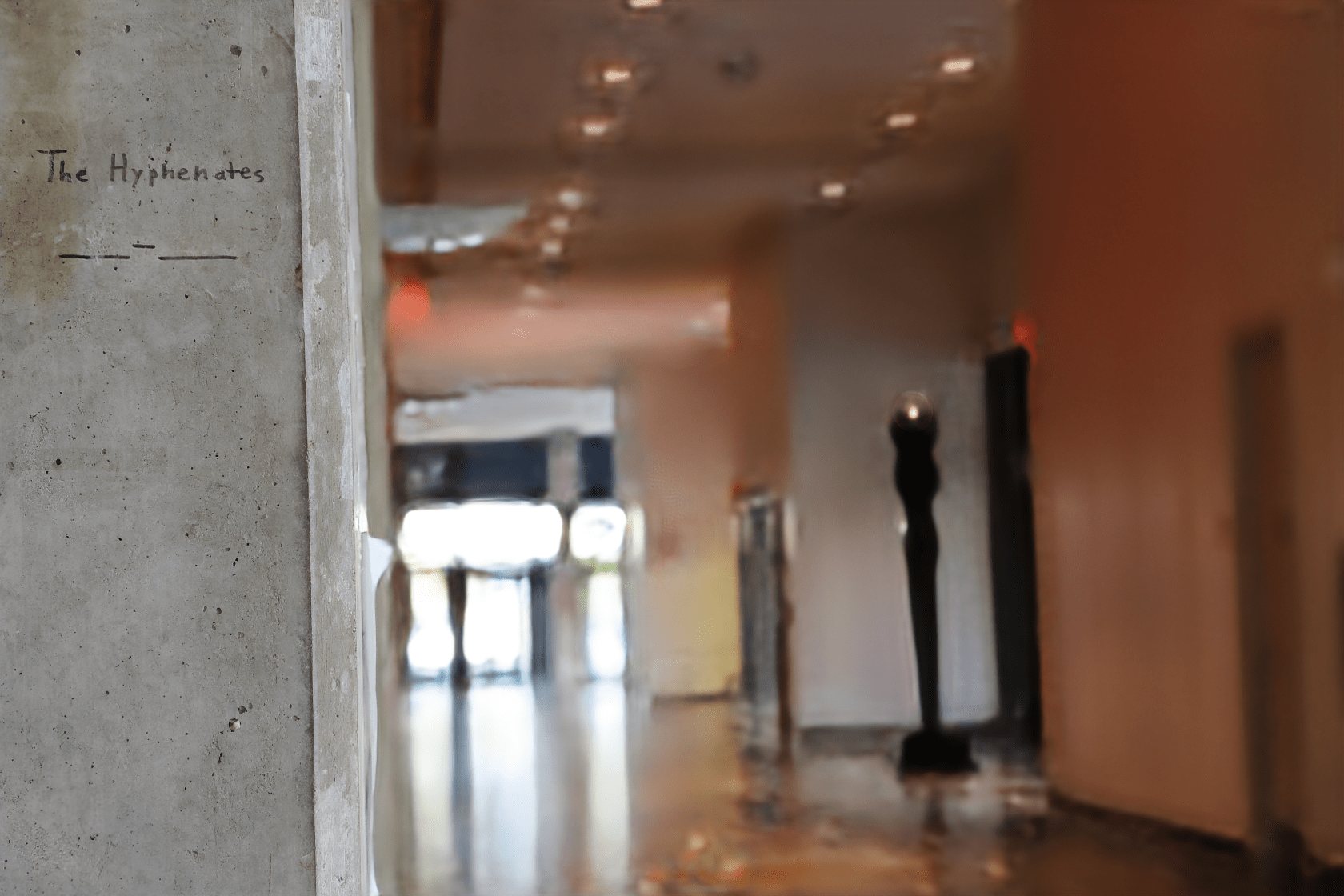}
        \end{subfigure}%
        \hspace{\fill}
        \begin{subfigure}[b]{0.245\textwidth}
                \includegraphics[width=\linewidth]{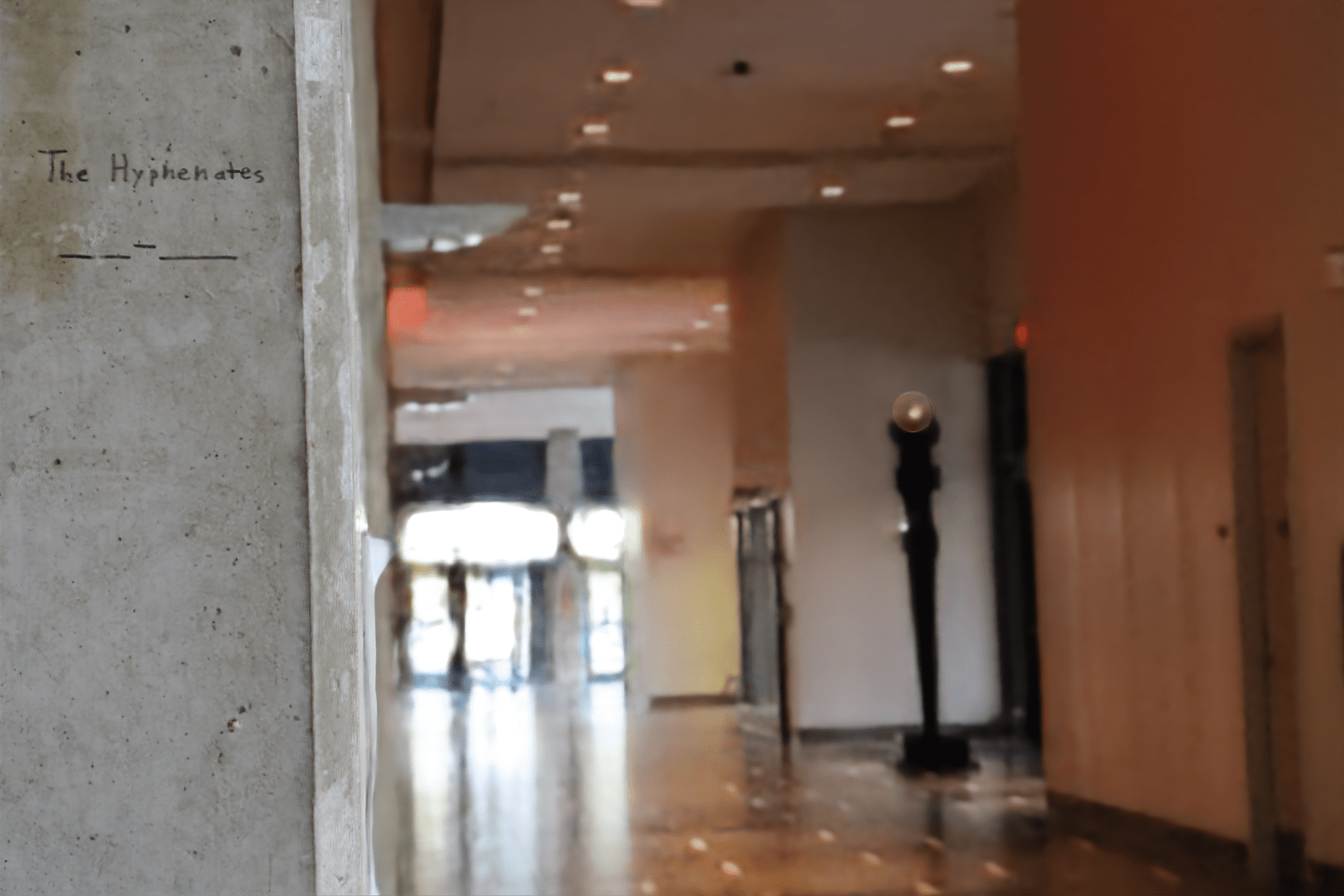}
        \end{subfigure}

        \begin{subfigure}[b]{0.245\textwidth}
                \includegraphics[width=\linewidth]{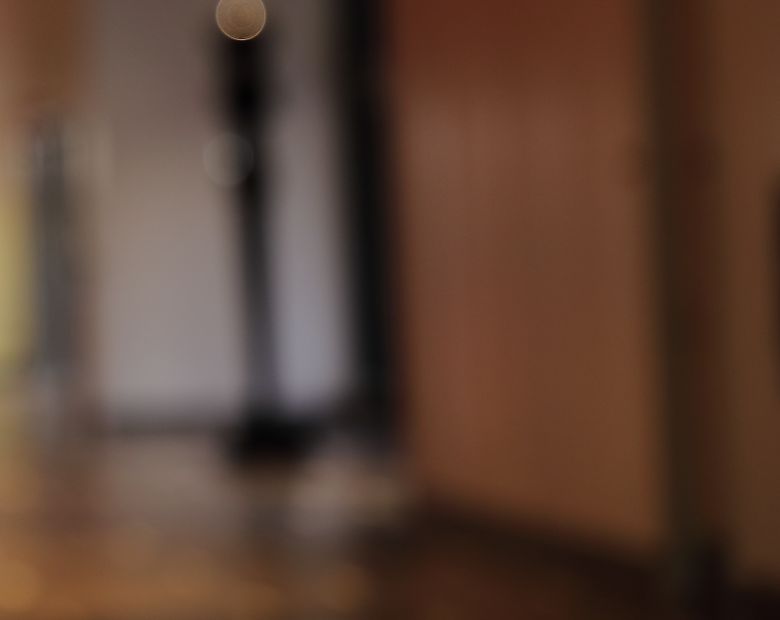}
                \caption{Input}
                \label{fig:gull}
        \end{subfigure}%
        \hspace{\fill}
        \begin{subfigure}[b]{0.245\textwidth}
                \includegraphics[width=\linewidth]{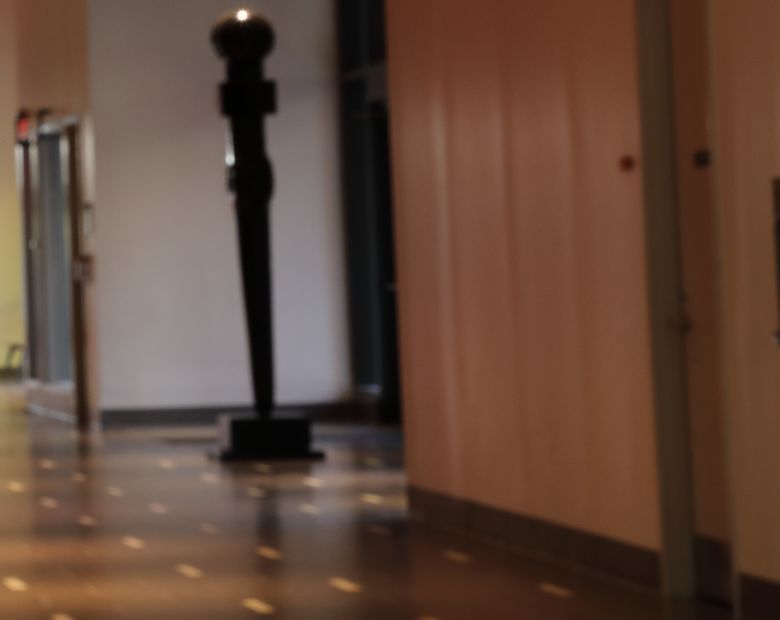}
                \caption{Ground Truth}
                \label{fig:gull2}
        \end{subfigure}%
        \hspace{\fill}
        \begin{subfigure}[b]{0.245\textwidth}
                \includegraphics[width=\linewidth]{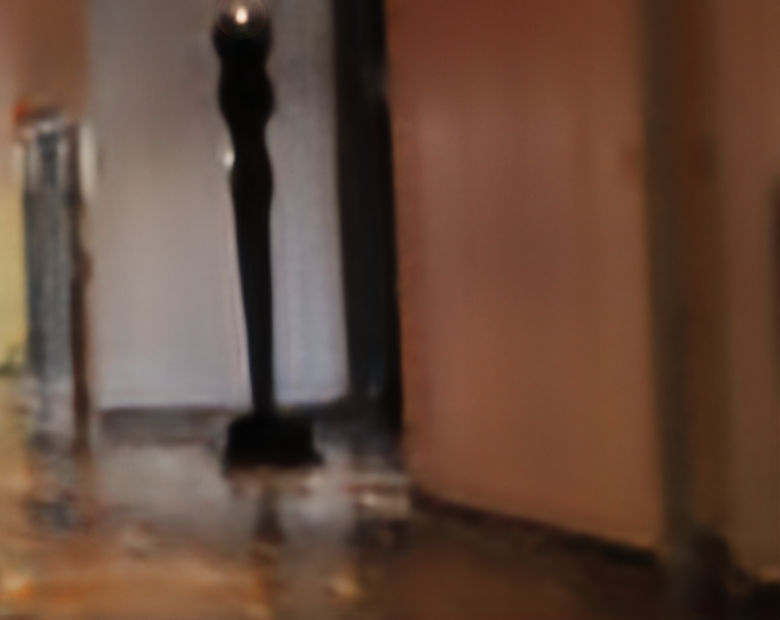}
                \caption{Abuolaim \textit{et al.}~\cite{abuolaim2020defocus} }
                \label{fig:tiger}
        \end{subfigure}%
        \hspace{\fill}
        \begin{subfigure}[b]{0.245\textwidth}
                \includegraphics[width=\linewidth]{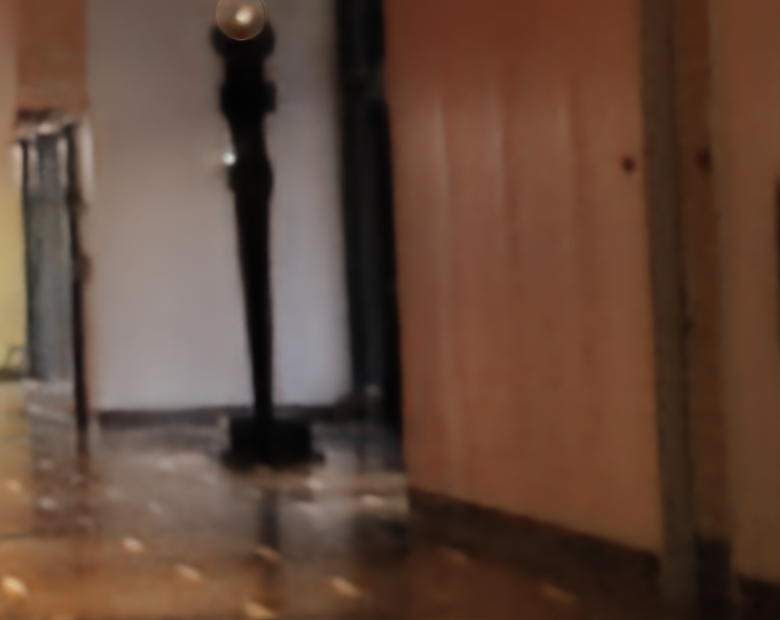}
                \caption{Proposed}
                \label{fig:mouse}
        \end{subfigure}
        \caption{Visual comparison of the proposed algorithm and Abuolaim \textit{et al.}~\cite{abuolaim2020defocus}'s algorithm. The proposed algorithm successfully deblur the blurry regions such as the door or the light on the ceiling.}\label{fig:two}
\end{figure*}

\begin{figure*}

	\begin{subfigure}[b]{0.32\textwidth}
                \includegraphics[width=\linewidth]{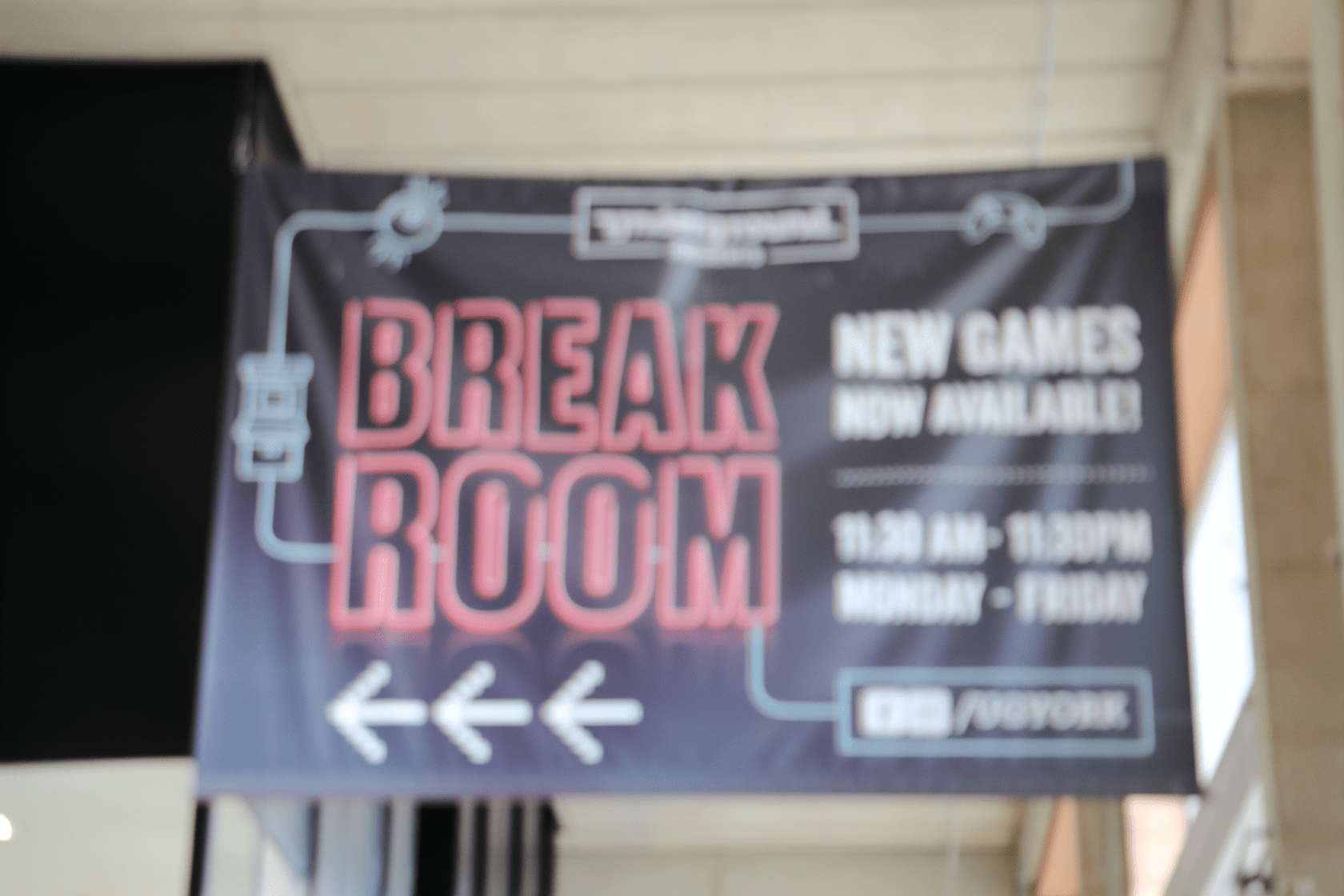}
        \end{subfigure}%
        \hspace{\fill}
        \begin{subfigure}[b]{0.32\textwidth}
                \includegraphics[width=\linewidth]{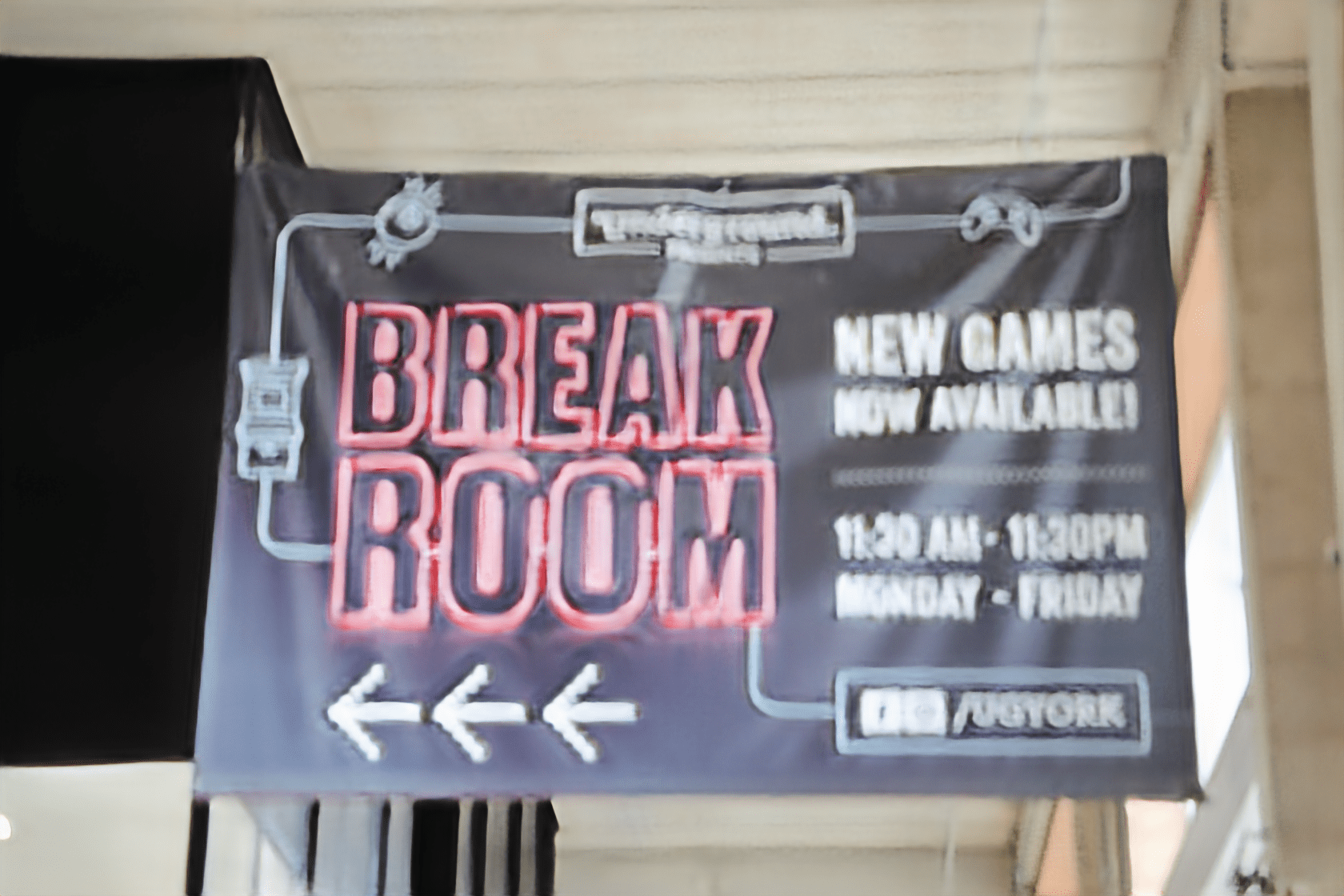}
        \end{subfigure}%
        \hspace{\fill}
        \begin{subfigure}[b]{0.32\textwidth}
                \includegraphics[width=\linewidth]{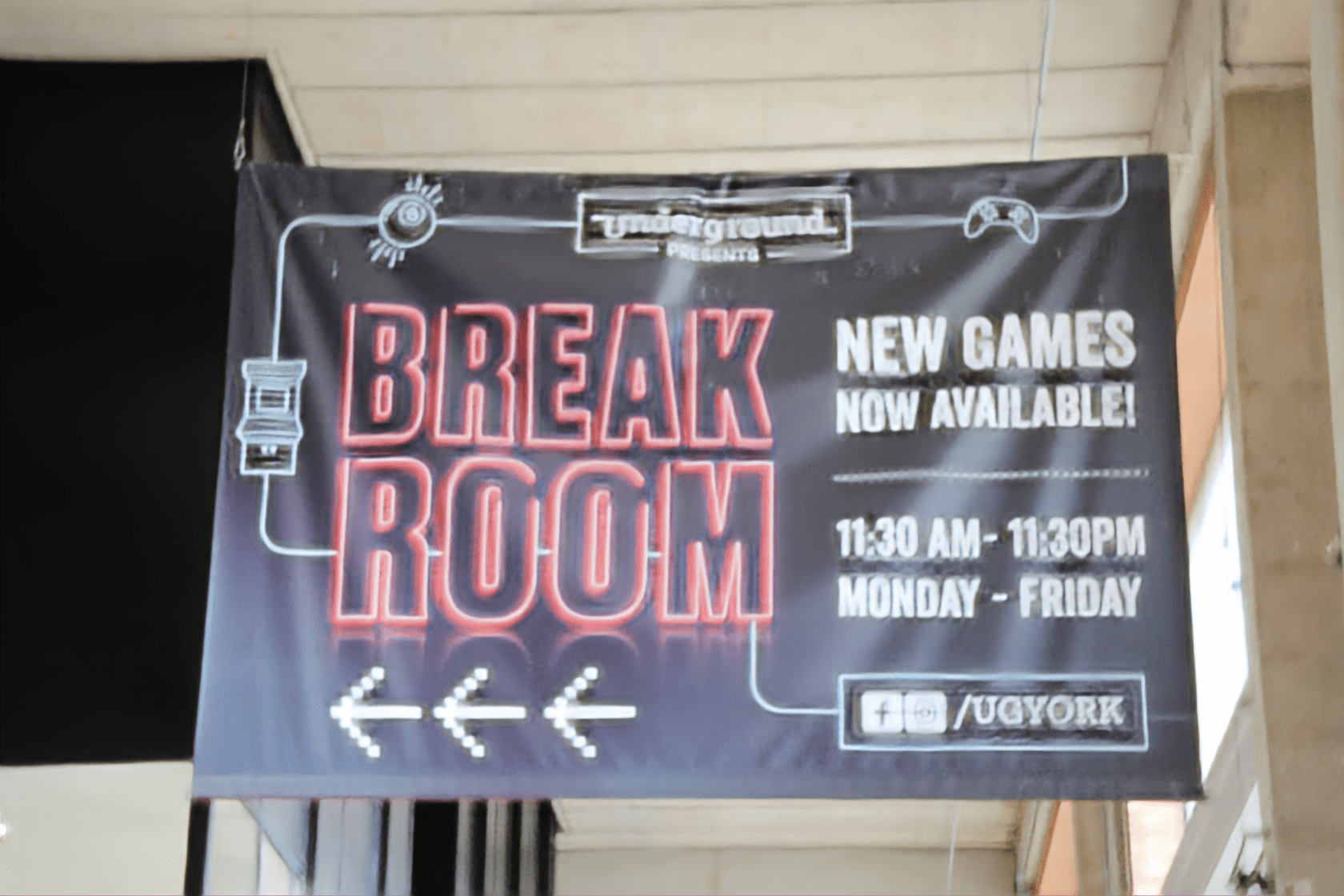}
        \end{subfigure}%
        \hspace{\fill}
        
        \begin{subfigure}[b]{0.32\textwidth}
                \includegraphics[width=\linewidth]{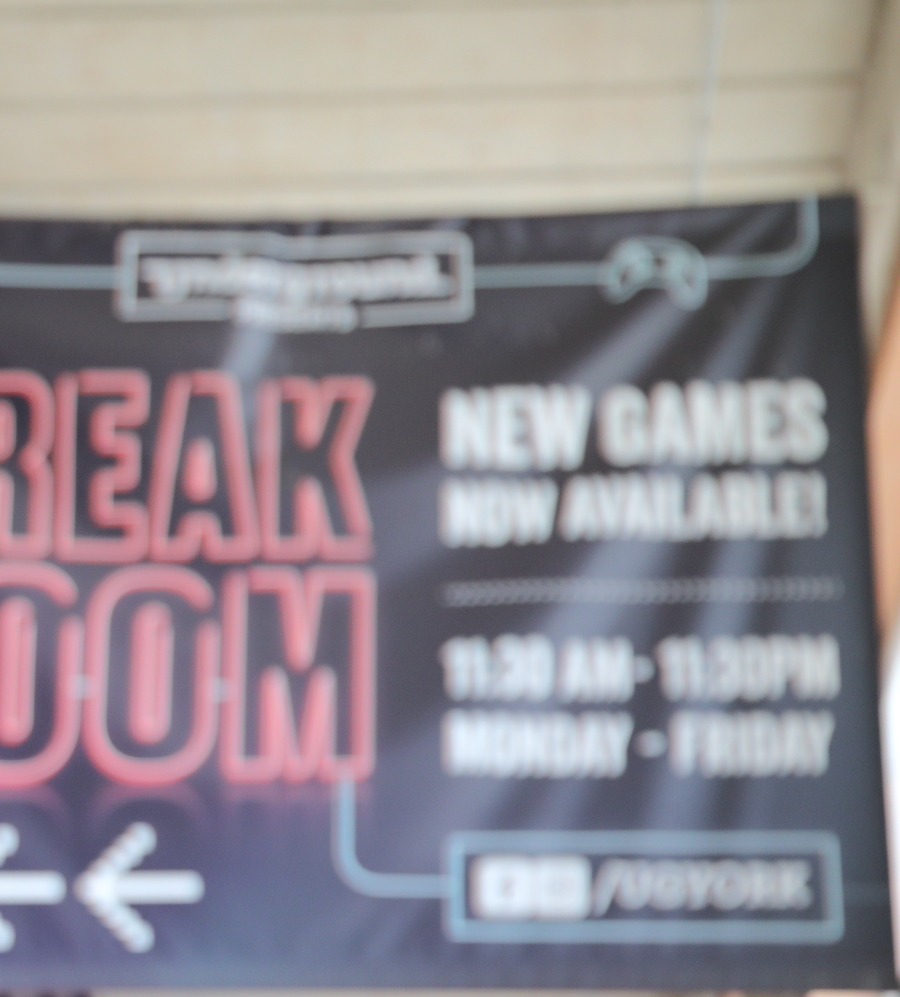}
        \end{subfigure}%
        \hspace{\fill}
        \begin{subfigure}[b]{0.32\textwidth}
                \includegraphics[width=\linewidth]{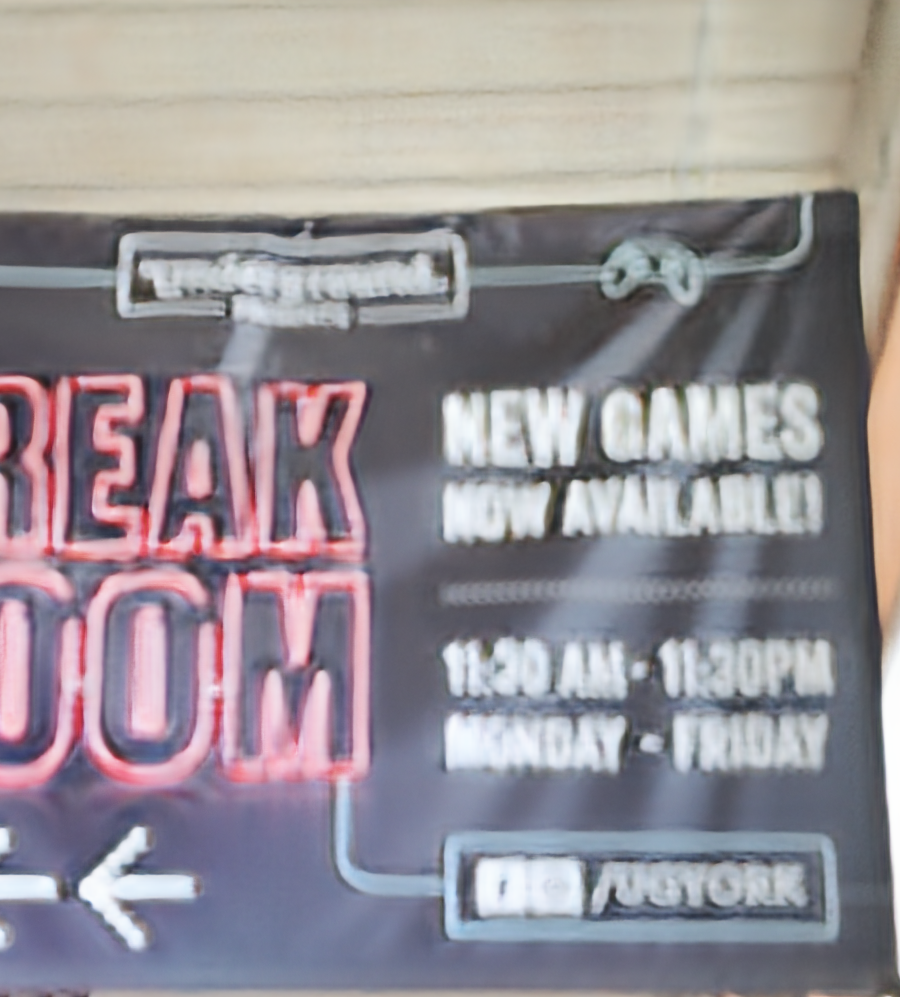}
        \end{subfigure}%
        \hspace{\fill}
        \begin{subfigure}[b]{0.32\textwidth}
                \includegraphics[width=\linewidth]{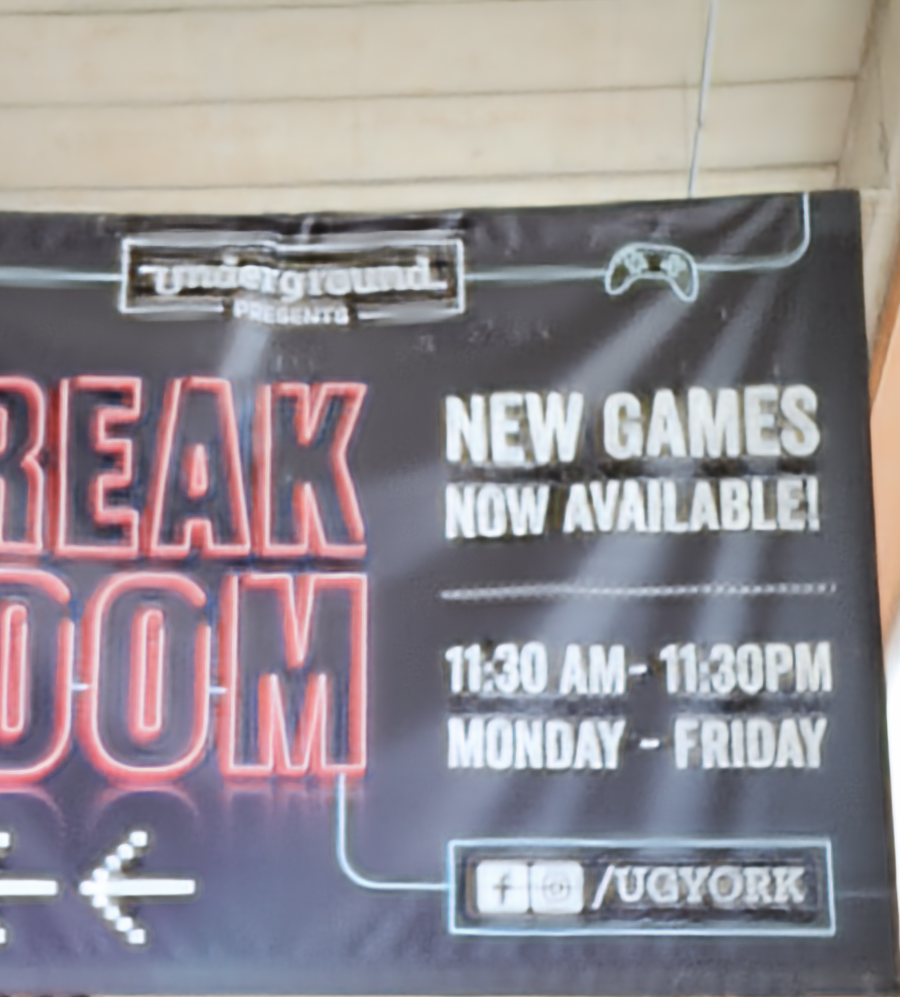}
        \end{subfigure}%
        \hspace{\fill}
	
        \begin{subfigure}[b]{0.32\textwidth}
                \includegraphics[width=\linewidth]{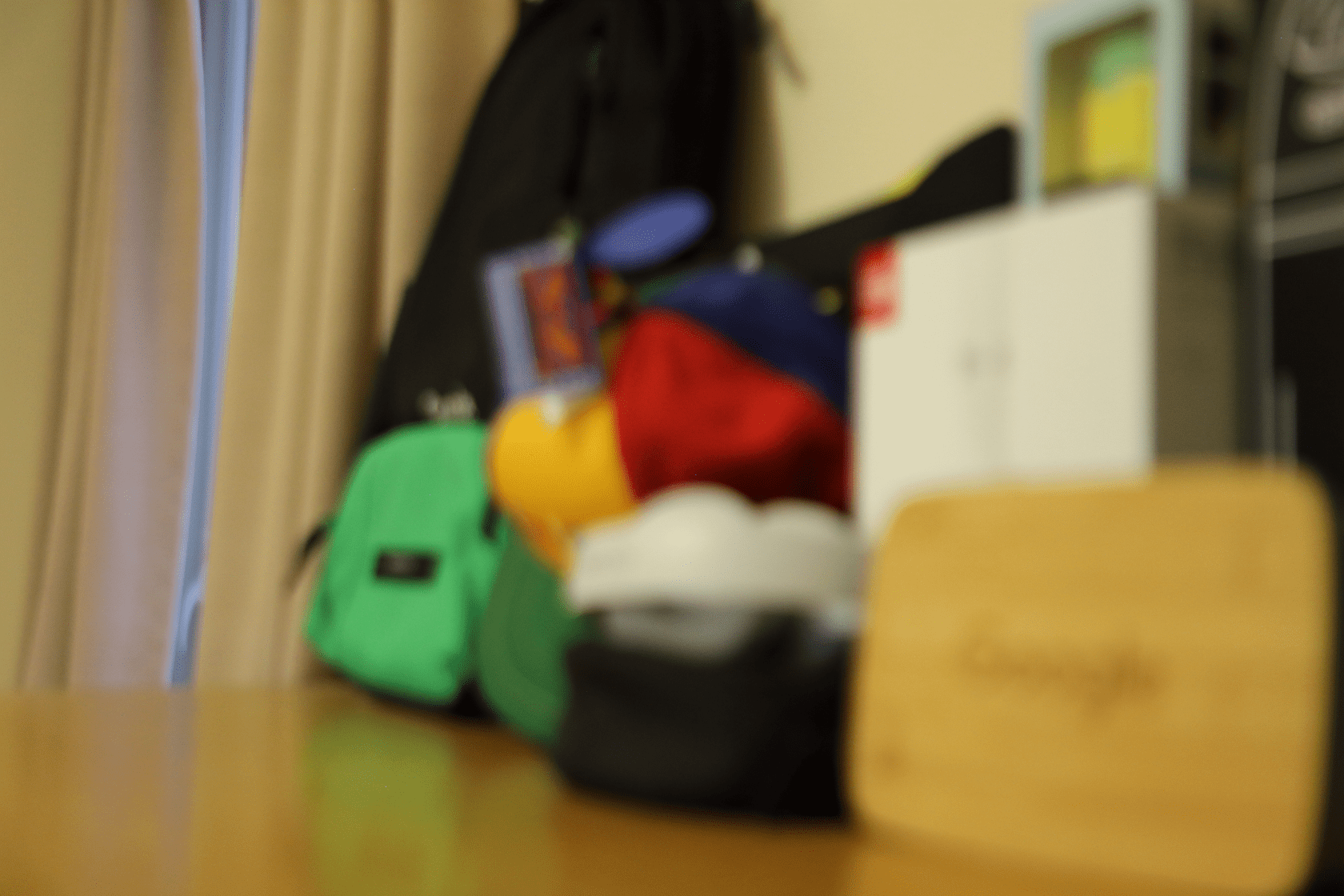}
        \end{subfigure}%
        \hspace{\fill}
        \begin{subfigure}[b]{0.32\textwidth}
                \includegraphics[width=\linewidth]{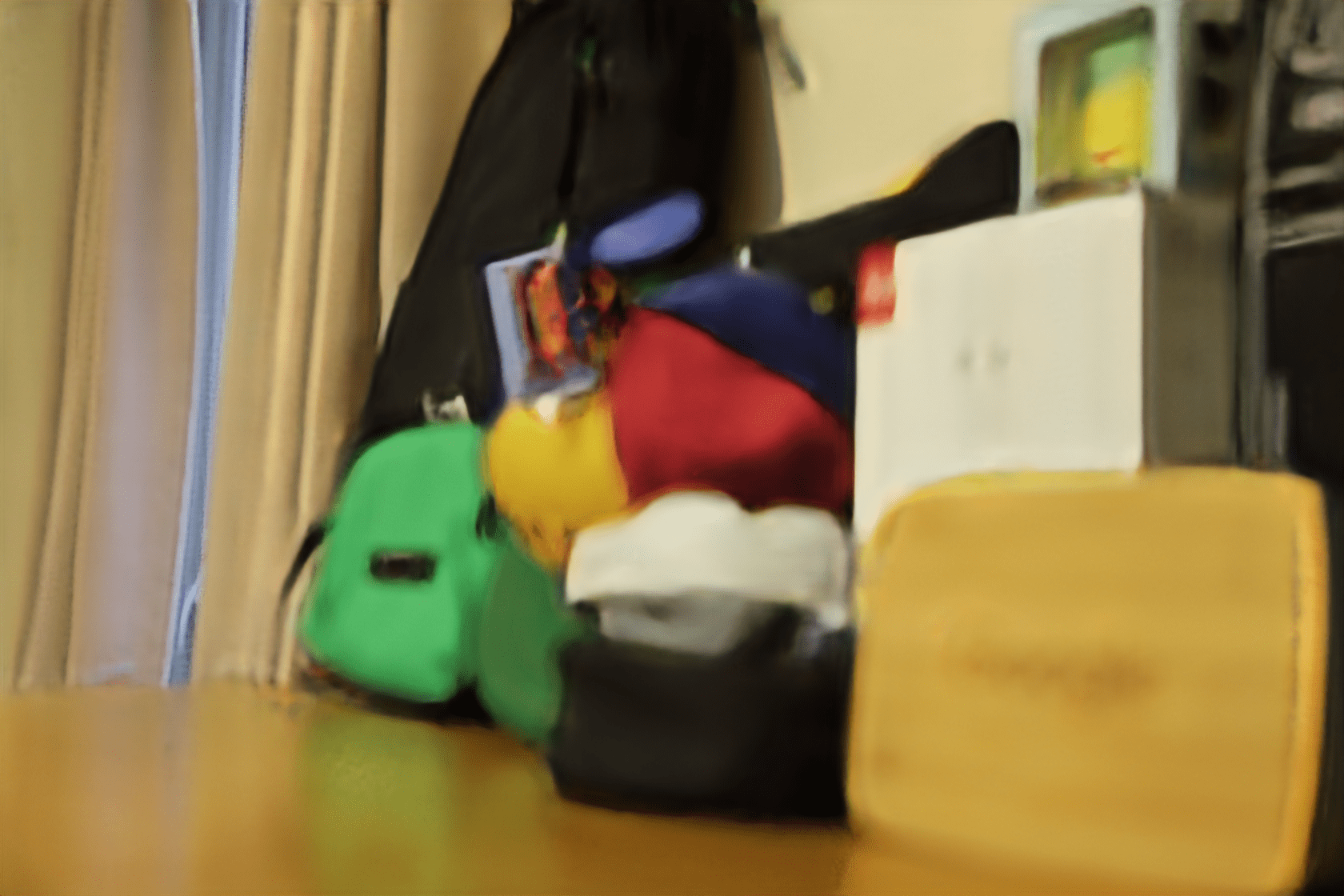}
        \end{subfigure}%
        \hspace{\fill}
        \begin{subfigure}[b]{0.32\textwidth}
                \includegraphics[width=\linewidth]{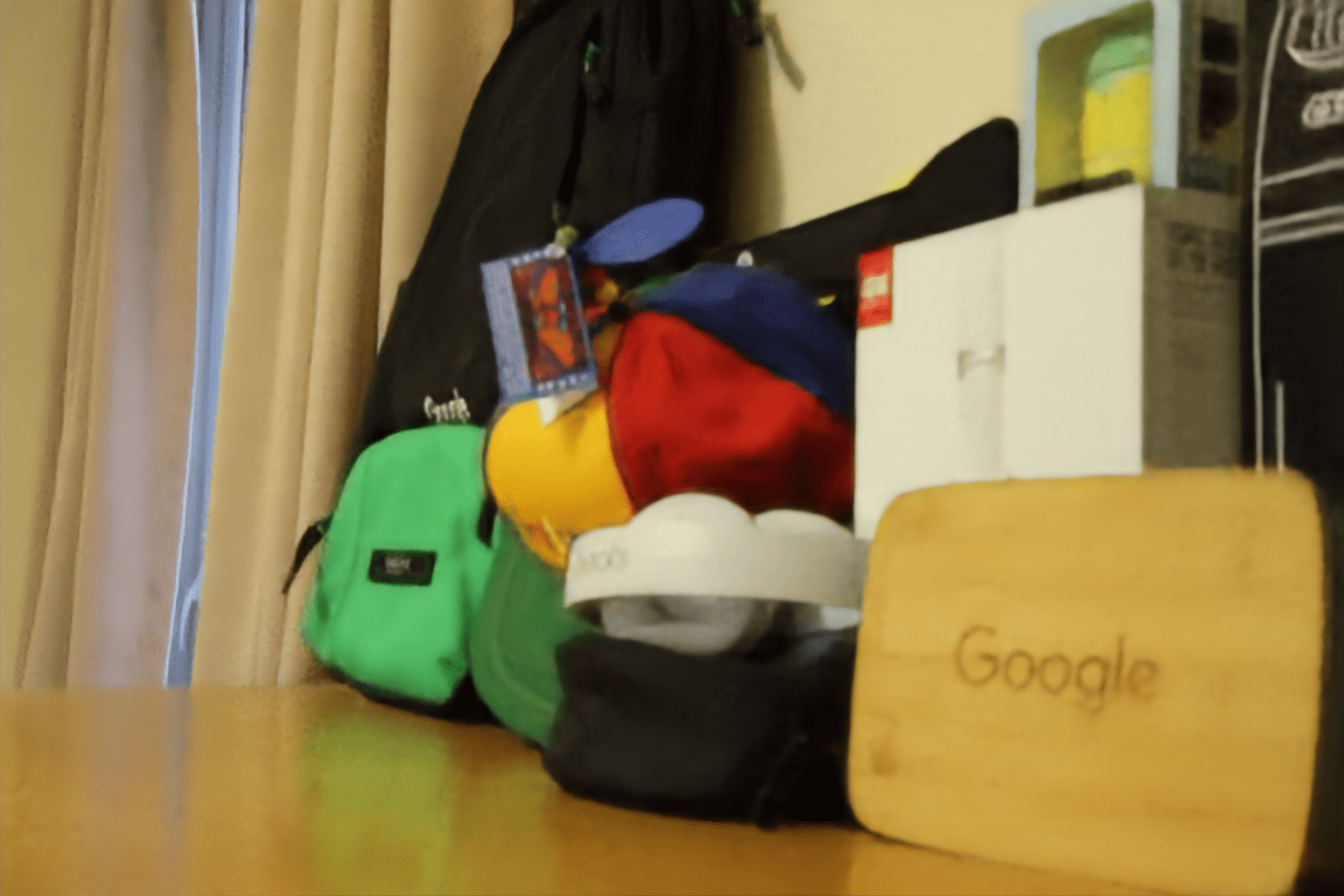}
        \end{subfigure}%
        \hspace{\fill}
        
        \begin{subfigure}[b]{0.32\textwidth}
                \includegraphics[width=\linewidth]{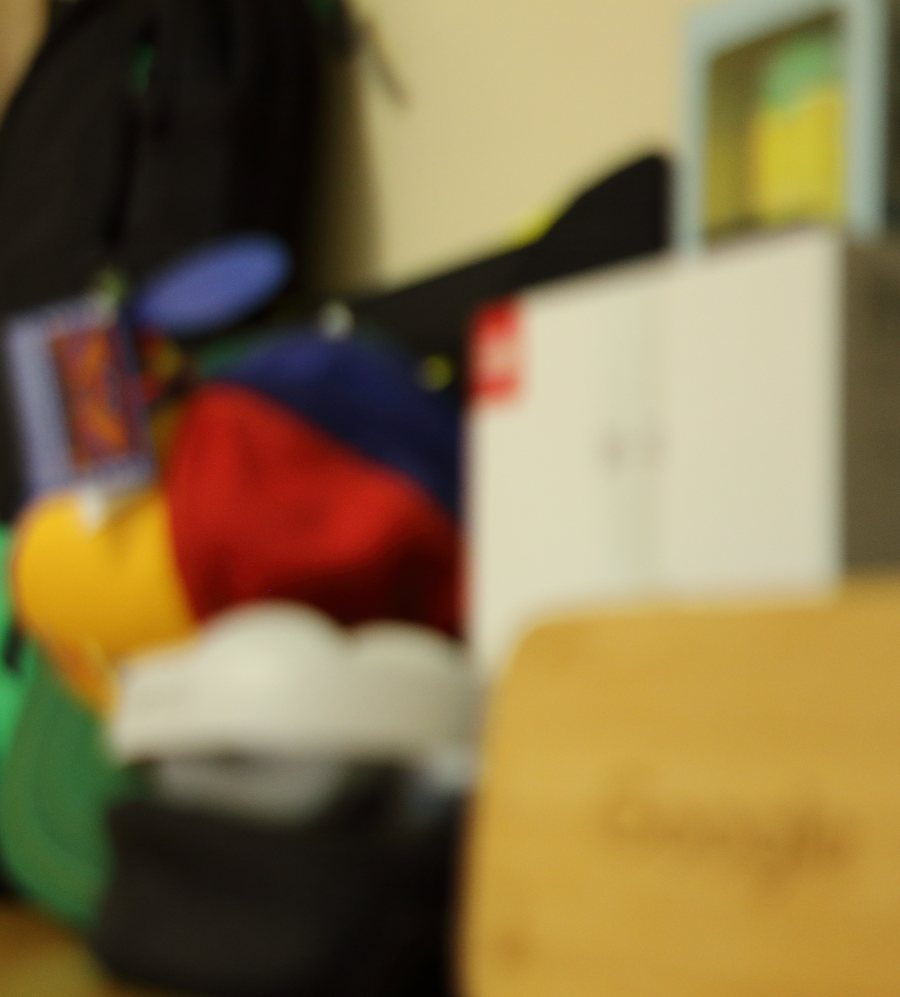}
                \caption{Input}
        \end{subfigure}%
        \hspace{\fill}
        \begin{subfigure}[b]{0.32\textwidth}
                \includegraphics[width=\linewidth]{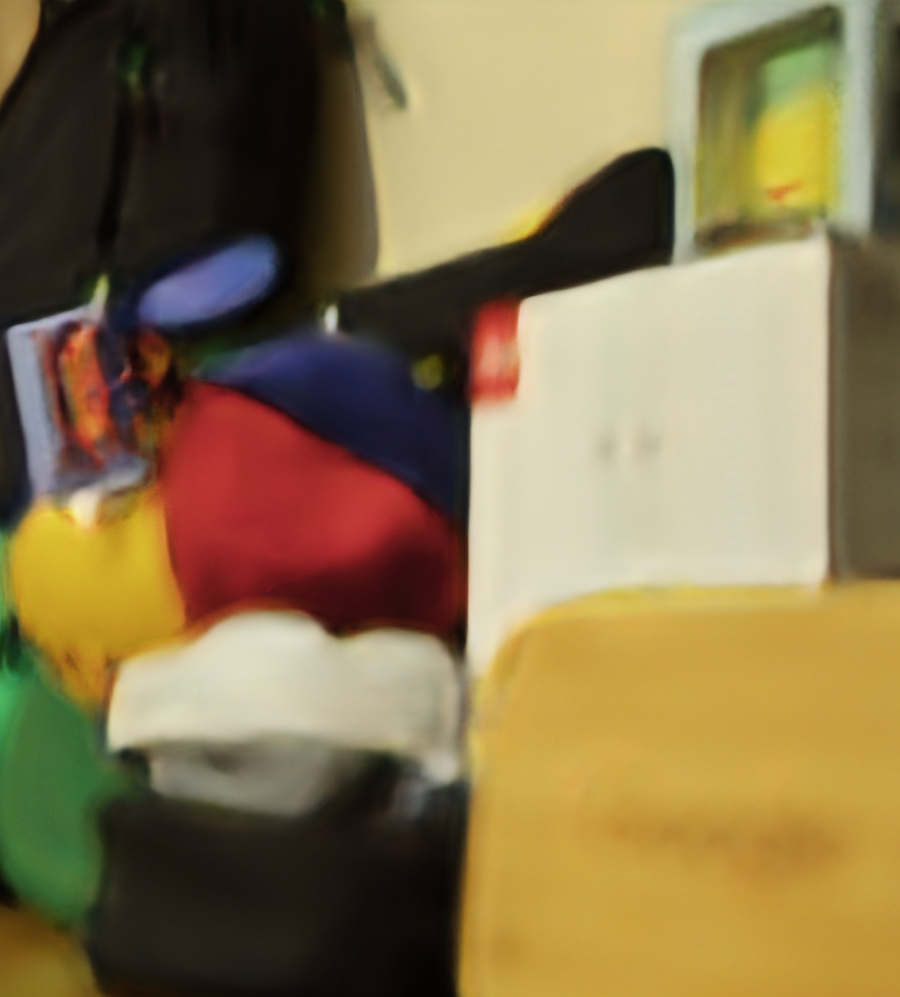}
                \caption{Abuolaim \textit{et al.}~\cite{abuolaim2020defocus} }
        \end{subfigure}%
        \hspace{\fill}
        \begin{subfigure}[b]{0.32\textwidth}
                \includegraphics[width=\linewidth]{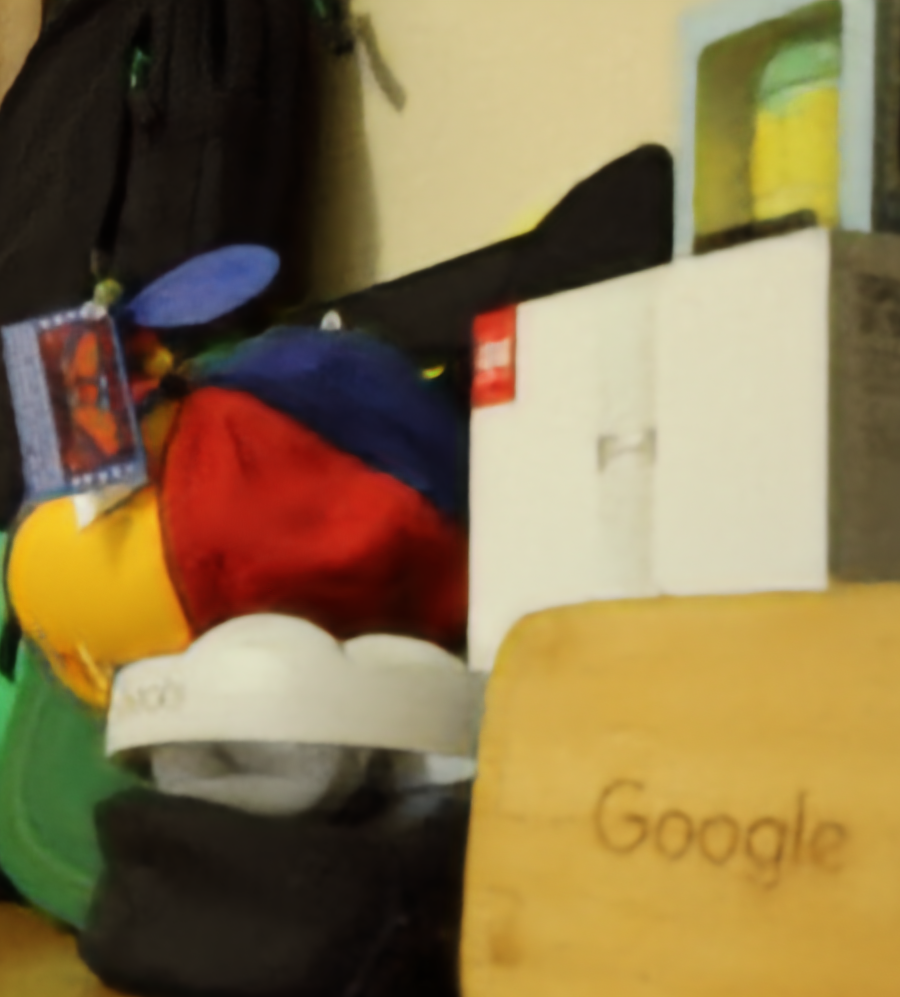}
                 \caption{Proposed}
        \label{fig:visual}
        \end{subfigure}%
        \hspace{\fill}
       
        \caption{Visual comparison of the proposed algorithm and Abuolaim \textit{et al.}~\cite{abuolaim2020defocus}'s algorithm. Abuolaim \textit{et al.}~\cite{abuolaim2020defocus} fails to produce the non-blur output while the proposed algorithm faithfully remove the blur artifact and generate sharp images.}\label{fig:three}
\end{figure*}

Figure~\ref{fig:one} and Figure~\ref{fig:two} visually compare the defocus deblurring results on the test set with ground truths provided. Abuolaim \textit{et al.}~\cite{abuolaim2020defocus} still yields blur artifacts while ATTSF preserves sharp edges and fine details more faithfully. We also verify the effectiveness of the our proposed method using the second test set provided by the competition. As this set does not contain the ground truth images, we only able to compare our results without the ground truth. The results shown on Figure~\ref{fig:three} again show that our proposed method successfully recover the blur region and out-perform the state-of-the-art algorithm. Although there was no ground truth to compare the results qualitatively, it is reported that the our \textit{PSNR} on this test set was $26.4243$ \textit{dB}, \nth{9} position in the competition.

\section{Conclusion}
In this work, we proposed an attention deep learning network which leverages the original encoder and decoder architecture by adding the dual-attention modules before every encoder blocks to attentionally extract the feature in each blur input image. Furthermore, at the bottleneck point, we also added the triple local and global local modules in parallel to efficiently extract the local features in different level as well as keep the global context of the input images. The features are then being concatenated with the encoded feature at every level and being decoded by the decoder modules, then finally restore the sharp output image. We demonstrated the effectiveness of the proposed defocus deblurring architecture through the \textit{NTIRE2021 Defocus Deblurring Challenge~\cite{ntire}}.

\bibliographystyle{ieee_fullname}
\bibliography{reference}

\end{document}